\documentclass[12pt]{article}

\usepackage{latexsym}
\usepackage{amsmath}
\usepackage{amsfonts}
\usepackage{mathrsfs}
\usepackage{dsfont}
\usepackage{multibox}
\usepackage{verbatim}

 \csname
@addtoreset\endcsname{equation}{section}

\def\g{\gamma}
\def\G{\Gamma}
\def\d{\delta}

\def\e{\epsilon}
\def\ve{\varepsilon}

\def\h{\eta}
\def\th{\theta}

\def\l{\lambda}
\def\L{\Lambda}
\def\m{\mu}
\def\n{\nu}
\def\x{\xi}

\def\p{\pi}

\def\r{\rho}

\def\S{\Sigma}

\def\vf{\varphi}
\def\c{\chi}

\def\o{\omega}

\def\cF{{\cal F}}

\def\cL{{\cal L}}
\def\cM{{\cal M}}
\def\cN{{\cal N}}
\def\cO{{\cal O}}
\def\cP{{\cal P}}

\def\cT{{\cal T}}

\def\cW{{\cal W}}

\def\be{\begin{equation}}
\def\ee{\end{equation}}
\def\bea{\begin{eqnarray}}
\def\eea{\end{eqnarray}}
\def\ba{\begin{array}}
\def\ea{\end{array}}

\def\nn{\nonumber}

\def\tr{\text{tr}}
\def\dim{3}

\def\ww{\wedge}

\def\comma{\,,\,}

\def\pe{\prime}
\def\12{\frac{1}{2}}
\def\pr{\partial}

\begin{document}

\begin{flushright}
AEI-2010-140
\end{flushright}

\vspace{25pt}

\begin{center}


{\Large\sc Asymptotic symmetries of three-dimensional gravity coupled
to higher-spin fields}


\vspace{25pt}
{\sc A.~Campoleoni, S.~Fredenhagen, S.~Pfenninger and S.~Theisen}

\vspace{10pt}
{\sl\small
Max-Planck-Institut f{\"u}r Gravitationsphysik\\
Albert-Einstein-Institut\\
Am M{\"u}hlenberg 1\\
14476 Golm,\ GERMANY\\[5pt]

\vspace{10pt}

{\it andrea.campoleoni@aei.mpg.de, stefan.fredenhagen@aei.mpg.de,\\  stefan.pfenninger@aei.mpg.de, stefan.theisen@aei.mpg.de} 
}

\vspace{70pt} {\sc\large Abstract}\end{center}

We discuss the emergence of $\cW$-algebras as asymptotic symmetries of
higher-spin gauge theories coupled to three-dimensional Einstein
gravity with a negative cosmological constant.  We focus on models
involving a finite number of bosonic higher-spin fields, and
especially on the example provided by the coupling of a spin-3 field
to gravity. It is described by a $SL(3)\times SL(3)$ Chern-Simons
theory and its asymptotic symmetry algebra is given by two copies of
the classical $\cW_3$-algebra with central charge the one computed by
Brown and Henneaux in pure gravity with negative cosmological
constant.

\newpage


\tableofcontents

\section{Introduction}

The covariant description of the free propagation of massless
higher-spin particles in a four-dimensional flat background was
obtained long ago by Fronsdal~\cite{fronsdal} in terms of gauge
theories. However, it was soon realised that the coupling to gravity
of the free actions displays various pathologies.\footnote{For a review
of the old no-go arguments and of the more recent results on how
higher-spin interactions in $D\geq4$ can nevertheless be constructed,
we refer the reader to \cite{review_interactions}. Other reviews on
various aspects of higher-spin gauge theories can be found in
\cite{review_vasiliev,reviewsHS,reviews_extra}.} For instance,
Aragone and Deser \cite{no-go} showed the inconsistency of the minimal
coupling of higher-spin gauge fields to gravity. At its heart the
obstruction rests on the impossibility to preserve the invariance of
the free action under higher-spin gauge transformations at the
interacting level.  In fact, the gauge variation of the minimally
coupled higher-spin actions is proportional to the full Riemann
tensor. As such, it cannot be cancelled by any variation of the
Einstein-Hilbert action. The highest-spin which is allowed is $3/2$
which leads to supergravity. There the gauge variation of the
Rarita-Schwinger action is proportional to the Ricci tensor rather
than to the Riemann tensor, and this is one of the crucial conditions
allowing supersymmetry.

In three space-time dimensions the situation is very different: on the
one hand, Fronsdal's gauge fields with ``spin'' $s>1$ do not propagate any
local degree of freedom.\footnote{The notion of spin we are referring
to is not related to the labelling of the representations of the
little group, which becomes trivial for massless particles in
$D=3$. It is simply associated to the transformation properties under
Lorentz transformation of the \emph{fields} we are going to consider.}
On the other hand, in $D=3$ the Weyl tensor vanishes for any
gravitational background and this suggests a possible way to avoid the
no-go results for minimal coupling. This expectation was indeed
confirmed by Aragone and Deser in \cite{hypergravity}.  However, this
is not the unique example of consistent interactions for higher-spin
gauge fields: a long-term effort by Vasiliev provided an interacting
theory for an infinite tower of massless higher-spin fields in
constant curvature backgrounds of any dimension
\cite{vasiliev_int}. On the other hand, the three-dimensional
peculiarities offer interesting toy models without many of the
technical complications that emerge when dealing with higher-spin
fields in $D>3$. For instance, in \cite{hypergravity} it was realised
that in three dimensions there is no need to consider an infinite
number of higher-spin fields in order to obtain consistent
interactions.

For a long time this result was part of a collection of
three-dimensional curiosities, like the by now well-known result of
Brown and Henneaux \cite{BH} on the asymptotic symmetries of
three-dimensional pure gravity with a negative cosmological
constant. These authors first proved that -- when considering
asymptotically Anti-de Sitter spaces -- the group of asymptotic
symmetries is the conformal group in two dimensions. They also showed
that its canonical realisation in terms of Dirac brackets of global
charges possesses a central extension. While the emergence of the
conformal group in two dimensions can also be inferred from the
structure of the AdS conformal boundary, the latter observation was
rather unexpected. Later on the central charge identified by Brown and
Henneaux was shown to play a crucial role in possible microscopic
interpretations of the entropy of the BTZ black hole
\cite{BTZ,strominger}. More generally, in modern terms the
Brown-Henneaux results should be considered as precursors of the
$AdS_3/CFT_2$ correspondence.

In the present paper we extend the considerations of Brown and Henneaux from the pure spin-2 case to the more general setup were also fields with spin $s>2$ are present. We then show how the coupling of these higher-spin fields to three-dimensional gravity allows for an enhancement of the boundary conformal symmetry. In fact, in presence of a negative cosmological constant the coupled system displays in general an extended conformal symmetry acting on the space of asymptotically AdS solutions of the field equations. In three space-time dimensions Einstein gravity can thus be considered as the simplest example of a wide class of higher-spin gauge theories whose dynamics is described by a conformal field theory on the boundary. Even if we shall not deal with the details of the boundary theory, the results we are going to present set the stage for possible higher-spin extensions of the standard $AdS_3/CFT_2$ correspondence.

We analyse the structure of asymptotic symmetries mainly by focussing on the coupling of a spin-3 field to gravity with a negative cosmological constant. In this example the asymptotic symmetries are given by two copies of the classical Zamolodchikov $\cW_3$-algebra with a central charge 
\be \label{c_intro}
c \, = \, \frac{3\,l}{2\,G}
\ee
coinciding with the Brown-Henneaux one \cite{BH}. We also comment on
the relation between more general $\cW$-algebras (see~\cite{W} for a
review on $\cW$-algebras) and the asymptotic symmetries of higher-spin
gauge theories with richer spectra. Our discussion rests on another
important observation on the three-dimensional world: the option to
describe interactions of fields with spin $s>1$ by means of a
Chern-Simons (CS) action. This was realised in \cite{townsend1,witten} for
(super)gravity theories. In \cite{blencowe} Blencowe then proposed a
Chern-Simons action which describes an infinite tower of interacting
higher-spin fields. Blencowe's theory was then shown to belong to a
one-parameter family of ``topological'' higher-spin interacting
theories with unbounded spectra \cite{general3D}. This rich set of
gauge theories was further analysed by Vasiliev and collaborators (see
\cite{deformed_oscillator,vasiliev_3d} and references therein), who
also discussed their coupling to matter. On the other hand, as we
already pointed out, in three space-time dimensions there is no need
to consider an infinite tower of higher-spin fields to consistently
switch on interactions. The possibility of describing them via a CS
action is not a peculiarity of Blencowe's theory or its
generalisations. It also applies to gauge theories with a finite
number of higher-spin fields, and in particular to our spin-3 example.

In Section \ref{sec:HS} we show how to cast generic higher-spin gauge
theories in a Chern-Simons form. This requires a reformulation of the
higher-spin dynamics along the lines of the frame formalism of gravity
and the identification of a suitable gauge algebra. We discuss this
last point in detail in the spin-3 case, for which we single out the
$SL(3)\times SL(3)$ gauge group. We also show how this example fits in
the class of $SL(n)\times SL(n)$ CS theories, that describes
interactions between a group of fields where each integer spin between
$2$ and $n$ appears once. This formalism enables us to discuss
asymptotic symmetries as global symmetries of a CS theory subject to
proper boundary conditions. For this reason in Section
\ref{sec:boundary} we review some general results on CS theories on
manifolds with boundaries. They provide the basis of Section
\ref{sec:symmetries}, where we resume the spin-3 example. We identify
the precise set of boundary conditions which characterise the
asymptotically Anti-de Sitter solutions of the field equations. We
then derive the asymptotic symmetries which they imply and obtain a
centrally extended classical $\cW_3 \otimes \cW_3$
algebra.\footnote{See \cite{Bilal,deBoer} for other connections
between CS theories and $\cW$-algebras.} At the end of Section
\ref{sec:symmetries} we return to the metric-like formulation. First
we fix the relation between metric-like fields and their frame-like
counterparts at the non-linear level. Then we use this result to
translate the boundary conditions for the CS theory in terms of
fall-off conditions for the metric-like fields, thus enabling a more
direct comparison with the standard Brown-Henneaux results. Finally,
in Section \ref{sec:generic_group} we comment on the case of a general
gauge group by comparing our boundary conditions with those
implementing the Hamiltonian reduction of Wess-Zumino-Witten (WZW)
models to Toda theories~\cite{WZW-Toda}. In particular, we point out
the universal character of the Brown-Henneaux central charge
\eqref{c_intro}, that emerges in a wide class of higher-spin gauge
theories including the $SL(n)\times SL(n)$ example. Section
\ref{sec:conclusions} closes this paper with a brief summary of our
results. In two appendices we specify our conventions and collect some
useful formulae.

We have been informed by Marc Henneaux that he and Soo-Jong Rey have
also studied the issue of asymptotic symmetry algebras of higher-spin
theories in three dimensions \cite{henneaux_rey}.

\section{Coupling to gravity and the Chern-Simons action}\label{sec:HS}

In this section we first recall some standard facts on the free theory
for higher-spin gauge fields in $D \geq 4$. Recognising that the
structure of the field equations does not depend on $D$, we then
clarify our notion of higher-spin fields in $D=3$. Afterwards,
starting from the frame-like description of the free theory, we show
how in $D=3$ higher-spin gauge fields can be coupled to gravity via a
Chern-Simons action. We then provide a class of examples
by focussing on the coupling of a tower of fields of increasing spin
$2,3,\ldots,n$, that is described by a $SL(n)\times SL(n)$ CS
theory. We close this section describing in detail the simplest model
of this class, that features the coupling of a spin-3 gauge field to
gravity with a negative cosmological constant.

\subsection{Free theory}

In a Minkowski background of arbitrary dimension $D\geq4$ the free propagation of a bosonic massless spin-$s$ particle can be described via a fully symmetric 
rank-$s$ tensor $\vf_{\m_1 \ldots\, \m_s}$ satisfying the second-order field equation \cite{fronsdal}
\be \label{fronsdal}
\cF_{\m_1 \ldots\, \m_s} \, \equiv \, \Box \, \vf_{\m_1 \ldots\, \m_s} \, - \, \pr_{(\m_1|} \pr^{\,\l}\, \vf_{|\m_2 \ldots\, \m_s)\,\l} \, + \pr_{(\m_1}\pr_{\m_2} \vf_{\m_3 \ldots\, \m_s)\l}{}^\l \, = \, 0 \, ,
\ee
where here and in the following a pair of parentheses denotes a
complete symmetrisation of the indices it encloses, with the minimum
possible number of terms and without any normalisation factor. Eq.\ \eqref{fronsdal} is left invariant by the gauge transformation
\be \label{gauge_fronsdal}
\d\, \vf_{\m_1 \ldots\, \m_s} \, = \, \pr_{(\m_1} \x_{\,\m_2 \ldots\, \m_s)} 
\ee
with a traceless gauge parameter:
\be \label{traceL}
\x_{\,\m_1 \ldots\, \m_{s-3}\l}{}^\l \, = \, 0 \, .
\ee
Notice that in the spin-2 case $\cF_{\m\n}$ is the linearised Ricci tensor, and the transformation \eqref{gauge_fronsdal} is a linearised diffeomorphism. 
Imposing the double-trace constraint
\be \label{tracephi}
\vf_{\m_1 \ldots\, \m_{s-4}\l\r}{}^{\l\r} \, = \, 0
\ee
one can build a second-order Lagrangian that, up to total derivatives, is invariant under the constrained gauge transformations \eqref{gauge_fronsdal}. The resulting action, identified by Fronsdal \cite{fronsdal}, is
\be \label{lag_fronsdal}
S \,=\, \frac{1}{2} \int d^D x \ \vf^{\m_1 \ldots\, \m_s} \left(\, \cF_{\m_1 \ldots\, \m_s} -\,\frac{1}{2}\, \h_{(\m_1\m_2}\, \cF_{\m_3 \ldots\, \m_s)\,\l}{}^\l \,\right) \, ,
\ee
and leads to field equations that are equivalent to \eqref{fronsdal}. 
Gauge invariance fixes eq.\ \eqref{lag_fronsdal} up to a normalisation factor 
\cite{curtright}.
For more details see the reviews collected in \cite{reviewsHS}.

In an AdS background one can look for field equations which are invariant under the gauge transformations
\be \label{fronsdal_gauge}
\d \, \vf_{\m_1 \ldots\, \m_s} \, = \, \nabla_{(\m_1} \x_{\,\m_2 \ldots\, \m_s)} \, ,
\ee
where $\nabla_\m$ is the AdS covariant derivative. They describe the propagation of the same number of degrees of freedom as a massless spin-$s$ field in flat space \cite{fronsdal_ads}. However, covariant derivatives no longer commute: denoting by $l$ the AdS radius and by $g_{\m\n}$ the AdS metric one obtains
\be
\left[\, \nabla_\m \comma \nabla_\n \,\right]\, V_\r \, = \, \frac{1}{l^2}\, (\, g_{\n\r}\, V_\m \,-\, g_{\m\r}\, V_\n \,) \, .
\ee
Therefore, one has to add extra terms to the field equations in order to keep the gauge invariance. The result is
\be \label{fronsdalADS}
\cF_{\m_1 \ldots\, \m_s} - \, \frac{1}{l^2}\, \Big\{\, \left[\, s^2 + (D-6)s -2 (D-3) \,\right]\, \vf_{\m_1 \ldots\, \m_s} + \, 2 \,g_{(\m_1\m_2} \vf_{\m_3 \ldots\, \m_s)\,\l}{}^\l \,\Big\} \, = \, 0 \, .
\ee
Here $\cF$ denotes the combination entering eq.\ \eqref{fronsdal},
with the substitution $\pr_\m \to \nabla_\m$ \cite{fronsdal_ads}. Eq.\
\eqref{fronsdalADS} is gauge invariant only if the gauge parameter
satisfies eq.\ \eqref{traceL}. In order to relate these equations to
an action principle it is convenient to introduce the deformed
Fronsdal operator $\widehat{\cF}$, defined by the left-hand side of
eq.\ \eqref{fronsdalADS}. In fact, imposing the double-trace constraint \eqref{tracephi}, field equations equivalent to \eqref{fronsdalADS} follow from the action
\be \label{lag_fronsdal2}
S \,=\, \frac{1}{2} \int d^D x \sqrt{-g}\ \vf^{\m_1 \ldots\, \m_s} \left(\, \widehat{\cF}_{\m_1 \ldots\, \m_s} -\,\frac{1}{2}\, g_{(\m_1\m_2}\, \widehat{\cF}_{\m_3 \ldots\, \m_s)\,\l}{}^\l \,\right) \, .
\ee
Even in this case, its structure is fixed by the request of gauge invariance (up to a conventional ordering choice for the covariant derivatives) \cite{fronsdal_ads}.

In $D=\dim$ the little group of massless particles is the direct product of the multiplicative group $\{1,-1\}$ with $\mathbb{R}$ \cite{binegar}. As a result, excluding representations with continuous spin, one is left only with the two inequivalent representations of $\{1,-1\}$. The usual notion of spin in $D=\dim$ thus just reduces to a distinction between bosons and fermions \cite{binegar}. Nevertheless, one can still consider the field equations \eqref{fronsdal} or \eqref{fronsdalADS} for tensors of arbitrary rank. They force on-shell the propagation of a number of local degrees of freedom equal to the number of components of a traceless tensor of the same rank in $D-2$ dimensions \cite{reviewsHS}. Therefore, in $D=\dim$ they do not lead to the propagation of any local degree of freedom if the rank of the tensor is greater than one. However, even if the bulk dynamics is trivial, in presence of a cosmological constant fields with different rank lead to different boundary dynamics. This distinction motivates to denote as spin the rank $s$ of the field $\vf_{\m_1 \ldots\, \m_s}$. An higher-spin gauge field in $D=3$ is thus a fully symmetric field $\vf_{\m_1 \ldots\, \m_s}$ with $s>2$, that at the linearised level admits the gauge transformations \eqref{fronsdal_gauge} and satisfies the field equation \eqref{fronsdalADS}. 

Let us now present an alternative description of the free dynamics that will prove convenient to discuss interactions for these fields. As in the frame formulation of gravity one can substitute the fully symmetric Fronsdal field with a $1$-form $e_\m{}^{a_1 \ldots\, a_{s-1}}$ \cite{vasiliev0}. In the standard frame-like approach this vielbein-like field is traceless and fully symmetric in its flat indices. However, due to the curved index, it carries a hooked $\{s-1,1\}$ component\footnote{Here and in the following we classify tensors by representations of the permutation group acting on their indices. These are labelled by Young diagrams, that we identify by ordered lists of the lengths of their rows enclosed between braces.} that is absent in the fully symmetric $\vf_{\m_1 \ldots\, \m_s}$. To eliminate it one resorts to a local Lorentz-like gauge transformation with a $\{s-1,1\}$-projected parameter, so that
\be
\d\, e_\m{}^{a_1 \ldots\, a_{s-1}} \, = \, D_\m \, \x^{\,a_1 \ldots\, a_{s-1}} \, + \, \bar{e}_{\m\,,\,b}\, \L^{b\,,\,a_1 \ldots\, a_{s-1}} \, ,
\ee
where $D_\m$ is the Lorentz-covariant derivative while $\bar{e}_\m{}^a$ is the background vielbein. This leads to introduce a gauge connection $\o_\m{}^{b\,,\,a_1 \ldots\, a_{s-1}}$ for the new gauge transformation. It is the higher-spin analogue of the spin connection of gravity. The new field is traceless and $\{s-1,1\}$-projected in its flat indices. As in the gravity case, it must be an auxiliary field and it will be expressed in terms of $e_\m{}^{a_1 \ldots\, a_{s-1}}$ and its first derivatives via suitable torsion-like constraints.

In conclusion, a spin-$s$ field freely propagating in a constant curvature background of arbitrary dimension can be described by the pair of one-forms
\be \label{frame_fields}
e_\m{}^{a_1 \ldots\, a_{s-1}} \, , \qquad \o_\m{}^{b\,,\,a_1 \ldots\, a_{s-1}} \, ,
\ee
which are irreducible Lorentz tensors in the flat indices. The Fronsdal formulation is recovered by eliminating the auxiliary field $\o_\m{}^{b\,,\,a_1 \ldots\, a_{s-1}}$ and considering the Lorentz-like invariant combination
\be \label{metric_field}
\vf_{\m_1 \ldots\, \m_s} \, \equiv \, \frac{1}{s}\ \bar{e}_{(\m_1}{}^{a_1} \ldots\, \bar{e}_{\m_{s-1}}{}^{a_{s-1}}\, e_{\m_s)\, a_1 \ldots\, a_{s-1}} \, ,
\ee
with the gauge transformations induced by those of the vielbein-like potential. Notice that the tracelessness condition on the vielbein-like field $e_\m{}^{a_1 \ldots\, a_{s-1}}$ induces the Fronsdal double trace constraints \eqref{tracephi} on the metric-like field $\vf_{\m_1 \ldots\, \m_s}$. In \cite{vasiliev0} Vasiliev identified a first-order action for $e_\m{}^{a_1 \ldots\, a_{s-1}}$ and $\o_\m{}^{b\,,\,a_1 \ldots\, a_{s-1}}$ describing the correct spin-$s$ free dynamics in a four-dimensional Minkowski background. In \cite{vasiliev1} he extended this result to constant curvature spaces and to arbitrary space-time dimensions. An important observation is that in general the resulting action is invariant under an enlarged set of gauge transformations. For instance, in a Minkowski background the free action is left invariant by the transformations
\begin{align}
& \d\, e_\m{}^{a_1 \ldots\, a_{s-1}} \, = \, \pr_\m \, \x^{\,a_1 \ldots\, a_{s-1}} \, + \, \bar{e}_{\m\,,\,b}\, \L^{b\,,\,a_1 \ldots\, a_{s-1}} \, , \nn \\[5pt]
& \d\, \o_\m{}^{b\,,\,a_1 \ldots\, a_{s-1}} \, = \, \pr_\m \, \L^{b\,,\,a_1 \ldots\, a_{s-1}} \, + \, \bar{e}_{\m\,,\,c}\, \Theta^{\,bc\,,\,a_1 \ldots\, a_{s-1}} \, , \label{extra_stueck}
\end{align}
where $\Theta^{b_1b_2\,,\,a_1 \ldots\, a_{s-1}}$ is an additional
traceless and $\{s-1,2\}$-projected St{\"u}ckelberg-like gauge parameter. Similar expressions hold in the AdS case, for which we refer the interested reader to the second reference of \cite{vasiliev1}. The appearance of a new gauge parameter calls for the introduction of an extra gauge connection. The procedure iterates until all gauge connections
\be
\o_\m{}^{b_1\ldots\, b_t\,,\,a_1 \ldots\, a_{s-1}} \, , \qquad 2 \leq t \leq s-1 \, ,
\ee
are introduced. They are traceless and $\{s-1,t\}$-projected in their flat indices. They are usually called \emph{extra fields} and they are necessary in order to rewrite the field equations in terms of curvatures, i.e. in terms of relations between gauge invariant objects. Even if extra fields do not enter the free action they do play a crucial role in Vasiliev's interacting theory \cite{review_vasiliev}. However, in $D=\dim$ the gauge parameter $\Theta^{b_1b_2\,,\,a_1 \ldots\, a_{s-1}}$ vanishes and also the extra fields do. In fact, for $O(n)$ groups the representations associated to Young diagrams with more than $n$ boxes in the first two columns vanish (see for instance \cite{hamermesh}, \textsection 10-6). As a result, in the following we shall ignore extra fields.
Furthermore, restricting the attention to the proper orthogonal subgroups $O^+(n)$, the representations with $a$ boxes in the first column and those with $n-a$ boxes in the first column are equivalent (see again \cite{hamermesh}, \textsection 10-6). In the three-dimensional context an example of this fact is the possibility to use the connection
\be
\o_\m{}^a \, = \, \12\, \e^{abc}\, \o_{\m\,,\,b\,,\,c} \, ,
\ee
rather than the usual spin connection $\o_\m{}^{a\,,\,b}$. The same is thus true for generic higher spins. In $D=\dim$ they can be described by the pair of gauge potentials 
\be \label{potentials}
e_\m{}^{a_1 \ldots\, a_{s-1}} \, , \qquad \o_\m{}^{a_1 \ldots\, a_{s-1}} \, ,
\ee
sharing the same index structure. For more details on the frame-like formulation of the dynamics we refer to \cite{review_vasiliev} and the references therein. In the next section, we shall take this last observation as a starting point for extending to higher spins the Chern-Simons reformulation of three-dimensional gravity.

\subsection{Chern-Simons formulation}\label{sec:cs_lin}

In presence of a negative cosmological constant three-dimensional Einstein gravity is equivalent to a Chern-Simons theory with gauge group $SO(2,2) \sim SL(2,\mathbb{R}) \times SL(2,\mathbb{R})$ \cite{townsend1,witten}. A CS reformulation is available also for its supergravity extensions, with a gauge group which is the product of two supersymmetric extensions of $SL(2,\mathbb{R})$ \cite{townsend1}. In both cases the field equations are zero-curvature conditions and thus no local degrees of freedom are involved. As we discussed, in $D=\dim$ this property holds also for a gauge field $\vf_{\m_1 \ldots\, \m_s}$ satisfying the Fronsdal equation \eqref{fronsdalADS}, and in fact in \cite{blencowe} Blencowe proposed an interacting theory for higher-spin fields in $D=\dim$ based on a CS action. In particular, he considered a gauge group which is the product of two-copies of an infinite-dimensional extension of $SL(2,\mathbb{R})$, thus mimicking the Fradkin-Vasiliev algebra driving higher-spin interactions in a four-dimensional AdS background \cite{fradkin_vasiliev}. However, as repeatedly stressed in the Introduction, in $D=\dim$ there is no need to consider an infinite tower of higher-spin fields in order to obtain consistent interactions. Therefore, in the following we shall review Blencowe's idea identifying the basic structures needed to couple any given spin-$s$ gauge field to gravity. However, in general this could require the simultaneous presence of other fields with different spin. 

In order to reformulate Einstein gravity in $D=\dim$ as a CS theory, 
one defines linear combinations of dreibein and spin connection as 
($l$ denotes the AdS radius) 
\be \label{gravity_potentials}
\jmath_\m{}^{a} \, = \, \o_\m{}^a  +  \frac{1}{l} \, e_\m{}^a \, , \qquad \tilde{\jmath}_\m{}^{a} \, = \, \o_\m{}^a  -  \frac{1}{l} \, e_\m{}^a \, ,
\ee
and interprets $\jmath$ and $\tilde{\jmath}$ as $sl(2,\mathbb{R})$ gauge potentials. In a similar fashion one defines the linear combinations
\be \label{forms1}
t_\m{}^{a_1 \ldots\, a_{s-1}} \, = \, (\, \o + \frac{e}{l} \,){}_\m{}^{a_1 \ldots\, a_{s-1}} \, , \qquad \tilde{t}_\m{}^{a_1 \ldots\, a_{s-1}} \, = \, (\, \o  -  \frac{e}{l} \,){}_\m{}^{a_1 \ldots\, a_{s-1}} \, .
\ee
of the fields \eqref{potentials}.
One  contracts them with some higher-spin generators $T_{a_1 \ldots\, a_{s-1}}$, to be added to the $sl(2,\mathbb{R})$ ones, and considers the one-forms
\begin{align} 
& A \, = \, \big(\, \jmath_\m{}^a \, J_a \, + \, t_\m{}^{a_1 \ldots\, a_{s-1}} \, T_{a_1 \ldots\, a_{s-1}} \,\big) \, dx^\m\, , \nn \\[5pt]
& \widetilde{A} \, = \, \big(\, \tilde{\jmath}_\m{}^a \,J_a \, + \, \tilde{t}_\m{}^{a_1 \ldots\, a_{s-1}} \, T_{a_1 \ldots\, a_{s-1}} \,\big) \, dx^\m \, . \label{forms}
\end{align}
Since no local degrees of freedom should be involved, in $D=3$ it is natural to identify the equations of motion for a spin-$s$ gauge field coupled to gravity with flatness conditions for $A$ and $\widetilde{A}$. This leads, at the action level, to a CS theory. We shall now support this conclusion by checking that the resulting field equations reduce to the Fronsdal one \eqref{fronsdalADS} at the linearised level. To this end, we have to impose conditions on the higher-spin generators. First of all, since they are contracted with the potentials \eqref{forms1}, they must transform as irreducible $so(1,2) \sim sl(2,\mathbb{R})$ tensors. Therefore, they must be symmetric and traceless in their indices 
(i.e. $T^b{}_{b a_3\dots a_{s-1}}=0$) and, as the $J_a$ satisfy
\be \label{pre_JJ}
\left[\, J_a \comma J_b \,\right] \, = \, \e_{abc}\, J^c \, ,
\ee
they must satisfy
\be \label{pre_JT}
\left[\, J_a \comma T_{b_1 \ldots\, b_{s-1}} \,\right] \, = \, \e^m{}_{a(b_1} T_{b_2 \ldots\, b_{s-1})m} \, .
\ee
If the $J_a$ and the $T_{a_1 \ldots\, a_{s-1}}$ generate a Lie algebra $\mathfrak{g}$ admitting a non-degenerate bilinear form (denoted in the following by $\textrm{tr}$) 
one can then consider the CS action
\be \label{chern-simons}
S_{CS}[A] \, = \, \frac{k}{4 \p} \, \int \textrm{tr} \left(\, A \ww d A \,+\, \frac{2}{3}\, A \ww A \ww A \,\right)\, .
\ee
In \cite{witten} it was pointed out that the combination
\be \label{action}
S \, = \, S_{CS}[A] - S_{CS}[\widetilde{A}]
\ee
reduces to the Einstein-Hilbert action, up to boundary terms, when $A$ and $\widetilde{A}$ only contain the gravitational fields $\jmath$ and $\tilde{\jmath}$. In particular, with the conventional normalization
\be \label{convention_killing}
\textrm{tr}(J_aJ_b) \, = \, \frac{1}{2}\, \h_{ab}
\ee
this identification leads to the relation
\be \label{level}
k \, = \, \frac{l}{4 G} \, ,
\ee
where $G$ is Newton's constant. As a result, eq.\ \eqref{action} provides the correct description of the gravitational sector, and we can check that the linearisation of its equations of motion also describes the free-propagation of a spin-$s$ field $\vf_{\m_1 \ldots\, \m_s}$ on an $AdS_3$ background. This ensures that the full interacting theory describes the coupling of $\vf_{\m_1 \ldots\, \m_s}$ to gravity.

To linearise the field equations derived from the action \eqref{action}, one splits  the gravitational dreibein and spin connection into background, 
$\bar{e}_\m{}^a$ and $\bar{\o}_\m{}^a$, and fluctuations,
\be
e_\m{}^a \, = \, \bar{e}_\m{}^a \, + \, h_\m{}^a \, , \qquad\qquad \o_\m{}^a \, = \, \bar{\o}_\m{}^a \, + \, v_\m{}^a \, ,
\ee
and treats the  higher-spin fields as fluctuations around trivial background values.
Notice that the commutator of two higher-spin generators is not needed for the linearised field equations.
Returning to the description in terms of the potentials \eqref{potentials}, the commutators \eqref{pre_JJ} and \eqref{pre_JT} imply that the spin-2 fluctuations satisfy
\begin{align}
& \mathscr{T}^a \, \equiv \, D\, h^a \,+\, \e^{abc}\, \bar{e}_b \ww v_c  = \, 0 \, , \nn \\
& \mathscr{R}^a \, \equiv \, D\, v^a \,+\, \frac{1}{l^2}\, \e^{abc}\, \bar{e}_b \ww h_c \, = \, 0 \, , 
\end{align}
while the spin-$s$ fluctuations satisfy
\begin{align}
& \mathscr{T}^{a_1 \ldots\, a_{s-1}} \, \equiv \,  D\, h^{\,a_1 \ldots\, a_{s-1}} \,+\, \e^{cd(a_1|}\, \bar{e}_c \ww v_d{}^{|a_2 \ldots\, a_{s-1})} \, = \, 0 \, , \nn \\
& \mathscr{R}^{a_1 \ldots\, a_{s-1}} \, \equiv \,  D\, v^{\,a_1 \ldots\, a_{s-1}} \,+\, \frac{1}{l^2}\, \e^{cd(a_1|}\, \bar{e}_c \ww h_d{}^{|a_2 \ldots\, a_{s-1})} \, = \, 0 \, . \label{eq_lin}
\end{align}
For brevity we omitted the form indices and we introduced the AdS covariant exterior derivative
\be
D \, f^{\,a_1 \ldots\, a_n} \, = \, d \, f^{\,a_1 \ldots\, a_n} \, + \, \e^{cd(a_1|}\, \bar{\o}_c \ww f_d{}^{|a_2 \ldots\, a_{n})} \, .
\ee
Notice that the field equations for the graviton are a particular case of those for a generic higher-spin gauge field freely propagating in an $AdS_3$ background. 
These field equations are left invariant by the transformations
\begin{align}
& \d\, h^{\,a_1 \ldots\, a_{s-1}} \, = \, D\, \x^{\,a_1 \ldots\, a_{s-1}} \,+\, \e^{cd(a_1|}\, \bar{e}_c\, \L_d{}^{|a_2\ldots\,a_{s-1})} \, , \nn \\[5pt] 
& \d\, v^{\,a_1 \ldots\, a_{s-1}} \, = \, D\, \L^{a_1 \ldots\, a_{s-1}} \,+\, \frac{1}{l^2}\, \e^{cd(a_1|}\, \bar{e}_c\, \x_{\,d}{}^{|a_2 \ldots\,a_{s-1})} \, , \label{gauge_frame}
\end{align}
since their gauge variation is proportional to the field equations for
the background fields. This is the full set of gauge transformations
of the linearised action, confirming the absence of extra fields
in $D=3$.

We have thus identified the linearised field equations implied by the
CS action \eqref{action} associated to any Lie algebra $\mathfrak{g}
\oplus \mathfrak{g}$ with a semisimple $\mathfrak{g}$ generated by
$J_a$ and $T_{a_1 \ldots\, a_{s-1}}$ satisfying eqs.\ \eqref{pre_JJ}
and \eqref{pre_JT}. We can now verify that they imply the Fronsdal
equation \eqref{fronsdalADS} for the field $\vf_{\m_1 \ldots\, \m_s}$
of eq.\ \eqref{metric_field}, while the gauge transformations \eqref{gauge_frame} imply the gauge transformations \eqref{fronsdal_gauge} for $\vf_{\m_1 \ldots\, \m_s}$. After this last step we shall eventually present a class of simple Lie algebras fitting into this scheme.

We start by noticing that the first of eqs.\ \eqref{eq_lin} is a generalisation of the torsion constraint of pure gravity and it can be used to express $v_\m{}^{a_1 \ldots\, a_{s-1}}$ in terms of $h_\m{}^{a_1 \ldots\, a_{s-1}}$. In fact
\be \label{torsion_eq}
\e^{\m\n\r}\, \mathscr{T}_{\m\n}{}^{a_1 \ldots\, a_{s-1}} \, = \, 0
\ee
describes a square system of algebraic equations for the various
components of $v_\m{}^{a_1 \ldots\, a_{s-1}}$. This property holds
only in $D=\dim$, while in higher space-time dimensions the mismatch
between the number of equations and the number of components of $v$ is
another evidence of the need for the extra St{\"u}ckelberg-like gauge
symmetry of eq.\ \eqref{extra_stueck}. Moreover, we can exhibit the
general solution of eq.\ \eqref{torsion_eq}, thus proving that its determinant is different from zero when the background dreibein is invertible. It reads
\be \label{sol_torsion}
\begin{split}
& (s-1)^2 \, v^{\,b\,,\,a_1 \ldots\, a_{s-1}} = \, (s-3)\, \bar{e}^{\,b\,(a_1|}\, v_c{}^{c\,|a_2 \ldots\, a_{s-2})} - \, 2\, \bar{e}^{\,(a_1a_2|}\, v_c{}^{bc\,|a_3 \ldots\, a_{s-2})} \\
& +\, (s-2)\, \e^{bcd}\, \bar{e}^{\,\m}{}_c \, \bar{e}^{\,\n,(a_1|}\, D_{[\m} h_{\n]\,d}{}^{|a_2 \ldots\, a_{s-1})} - \, \e^{cd(a_1|} \,  \bar{e}^{\,\m}{}_c \, \bar{e}^{\,\n,b}\, D_{[\m} h_{\n]\,d}{}^{|a_2 \ldots\, a_{s-1})} \\
& -\, \e^{cd(a_1|} \,  \bar{e}^{\,\m}{}_c \, \bar{e}^{\,\n,|a_2|}\, D_{[\m} h_{\n]\,d}{}^{|a_3 \ldots\, a_{s-1})\,b} \, ,
\end{split}
\ee
where the mixed trace of $v$ is 
\be \label{sol_trace}
v_b{}^{b\, a_1\ldots\, a_{s-2}} = \, \frac{1}{2s}\, \e^{bcd} \, \bar{e}^{\,\m}{}_c \, \bar{e}^{\,\n}{}_d \, D_{[\m} h_{\n]\,b}{}^{a_1 \ldots\, a_{s-2}} \, .
\ee
In eqs.\ \eqref{sol_torsion} and \eqref{sol_trace} the square brackets denote the antisymmetrisation of the indices they enclose, again with unit overall normalisation. As in the gravity case, the invertibility of the background dreibein plays a crucial role in the identification of the relation \eqref{sol_torsion}.

Substituting the solution of eq.\ \eqref{torsion_eq} in the linearised CS action one then obtains a second-order action depending on 
\be
h_{\m_1 ,\, \m_2 \ldots\, \m_s} \, = \, \bar{e}_{\m_2}{}^{a_1} \ldots\, \bar{e}_{\m_{s}}{}^{a_{s-1}}\, h_{\m_1,\, a_1 \ldots\, a_{s-1}} \, .
\ee
But acting with the Lorentz-like gauge transformation
\eqref{gauge_frame} generated by $\L^{a_1 \ldots\, a_{s-1}}$ it is
possible to eliminate the $\{s,1\}$-component carried by this
combination. As a result, the action eventually depends only on the
field \eqref{metric_field}, whose gauge transformations can be deduced
by acting with \eqref{gauge_frame} in eq.\ \eqref{metric_field}. In performing this substitution one can also eliminate the background spin connection appearing in the result by using the vielbein postulate
\be \label{viel_post}
\pr_\m \, e_\n{}^a \, + \, \e^{\,a}{}_{\!bc}\, \o_\m{}^b e_\n{}^c \, - \, \G^\l{}_{\m\n} \, e_\l{}^a \, = \, 0 \, .
\ee
The $AdS_3$ Christoffel symbols so introduced enable one to cast the gauge transformation in the form
\be \label{gauge_metric}
\d\, \vf_{\m_1 \ldots\, \m_s} \, = \, \nabla_{(\m_1} \x_{\m_2 \ldots\, \m_s)} \, 
\ee
with
\be \label{par_lin}
\x_{\m_1 \ldots\, \m_{s-1}} \, = \, \bar{e}_{\m_1}{}^{a_1} \ldots \bar{e}_{\m_{s-1}}{}^{a_{s-1}} \, \x_{\,a_1 \ldots\, a_{s-1}} \, .
\ee
Moreover, the tracelessness conditions on $\x^{a_1 \ldots\, a_{s-1}}$ and $h_\m{}^{a_1 \ldots\, a_{s-1}}$ induce the Fronsdal constraints \eqref{traceL} and \eqref{tracephi} on $\x_{\m_1 \ldots\, \m_{s-1}}$ and $\vf_{\m_1 \ldots\, \m_s}$. The resulting action thus coincides with the Fronsdal one \eqref{lag_fronsdal2}, since its structure is fixed by the requirement of gauge invariance under the transformations \eqref{gauge_metric}.

To summarise, we have reduced the problem of finding a consistent
gravitational coupling for a spin-$s$ field to the problem of finding
a semisimple Lie algebra whose generators can be split in $J_a$ and
$T_{a_1 \ldots\, a_s}$  satisfying eqs.\ \eqref{pre_JJ} and \eqref{pre_JT}. The Jacobi identities and the trace constraints could impose strong restrictions and a priori it could be necessary to simultaneously consider more higher-spin fields to fulfil them.\footnote{In principle even the choice $[\, T_{a_1 \ldots a_{s-1}} \comma T_{b_1 \ldots b_{s-1}} \,] \, = \, 0$ would be consistent, but the resulting algebra actually describes only the free-propagation of higher-spin gauge fields.} A direct constructive approach could then end up in a rather non-trivial task, but in \cite{harmonics,general3D} it was shown how to describe a generic $sl(n)$ algebra in terms of generators $T_{a_1 \ldots\, a_{s-1}}$ with $2\leq s \leq n$, traceless and fully symmetric in the indices they carry. This result provides a first interesting class of examples fitting into the previous discussion. 

The starting point is the observation that any symmetrised product of $sl(2,\mathbb{R})$ generators of the form
\be \label{trick}
T_{a_1 \ldots\, a_{s-1}} \, \sim \, J_{(a_1} \ldots J_{a_{s-1})}
\ee
satisfies the commutator \eqref{pre_JT} with $J_a$. Their traceless projections thus satisfy the properties that identify possible higher-spin generators, but in general the commutator between generators with spins $s_1$ and $s_2$ produces a new $T$ with spin
\be
|s_1-s_2|+1 \leq s_3 \leq s_1+s_2-1 \, ,
\ee
thus preventing the realisation of a finite-dimensional algebra.
However, if one considers a $n$-dimensional representation for the
$J_a$, the $T_{a_1 \ldots\, a_{s-1}}$ are $n \times n$ matrices. The
tracelessness condition in the $a_n$ indices then implies that they
are traceless matrices. Furthermore, the whole set of matrices
generated by the combinations \eqref{trick} with $s \leq n$ contains
\mbox{$n^2-1$} independent elements
\cite{harmonics,general3D}. Therefore, even if this argument does not
suffice to identify the precise form of the commutators between
higher-spin generators, it ensures that the first $n-1$ products
\eqref{trick} generate the $sl(n)$ algebra when one deals with a
$n$-dimensional representation of $sl(2,\mathbb{R})$. The particular real
form that one realises depends on the choice of the normalisation of
eq.\ \eqref{trick}, as we shall see in the next section in the spin-3
example. This presentation of $sl(n)$ implies that a $SL(n)
\times SL(n)$ CS theory can be interpreted as describing the coupling
of a tower of fields of increasing spin $2,3,\ldots,n$, where each
value of the spin appears only once. In the limit $n\to \infty$ the
present construction leads to the higher-spin gauge theory based on
the algebra of area-preserving diffeomorphisms on a two-dimensional
hyperboloid, which was discussed in detail in \cite{general3D}. In the
following we shall mainly examine the properties of this class of
higher-spin gauge theories -- and in particular of its simplest
example describing the coupling of a spin-3 field to gravity --
confining to Section \ref{sec:generic_group} some comments on more
general alternatives.

\subsection{The spin-3 example} \label{sec:algebra}

In the spin-3 case eqs.\ \eqref{pre_JJ} and \eqref{pre_JT} allow for
the introduction of a non-trivial commutator between the higher-spin
generators $T_{ab}$, which is uniquely fixed by the Jacobi identity up
to a normalisation constant $\sigma$. The resulting non-Abelian Lie algebra is
\begin{subequations}\label{sl3}
\begin{align}
& \left[\, J_a \comma J_b \,\right] \, = \, \e_{abc}\, J^c \, , \label{JJ} \\[5pt]
& \left[\, J_a \comma T_{bc} \,\right] \, = \, \e^m{}_{a(b} T_{c)m} \, , \label{JT} \\[5pt]
& \left[\, T_{ab} \comma T_{cd} \,\right] \, = \, \sigma \left(\, \h_{a(c} \e_{d)bm} + \, \h_{b(c} \e_{d)am} \,\right) J^m \, , \label{TT}
\end{align}
\end{subequations}
where the $T_{ab}$ are traceless and symmetric in $a,b$. Notice that
the right-hand side of eq.\ \eqref{JT} is traceless in the indices
$b,c$, while the right-hand side of eq.\ \eqref{TT} is traceless in $a,b$ and $c,d$. Imposing 
$T^a{}_a=0$ is thus consistent. The algebra \eqref{sl3} has the quadratic Casimir
\be \label{casimir}
C \, = \, J_a J^a \, - \, \frac{1}{2\,\sigma}\ T_{ab}\, T^{ab} \, ,
\ee
that enables to define a non-degenerate bilinear form on it. Actually, one can show  that \eqref{sl3} is isomorphic to $sl(3,\mathbb{C})$, and thus matches the corresponding algebra obtained from the construction discussed at the end of the previous section. One can build its fundamental representation by defining the $T_{ab}$ generators as
\be
T_{ab} \, = \, \sqrt{-\sigma} \,\left( J_{(a}J_{b)} - \, \frac{2}{3}\, \h_{ab}\, J_c J^c \right) \, ,
\ee
where the $J_a$ are the $sl(2,\mathbb{R})$ generators in the 3-dimensional representation. The sign of $\sigma$ selects one of the two non-compact real forms of $sl(3,\mathbb{C})$. In fact, the non-compact subalgebra $sl(2,\mathbb{R})$ rules out the compact real form, while a real rescaling of the generators $T_{ab}$ can modify the absolute value of $\sigma$ but not its sign. In particular, $\sigma>0$ corresponds to $su(1,2)$, while $\sigma < 0$ corresponds to $sl(3,\mathbb{R})$. Notice that the analogy with the spin-2 case is not sufficient to single out one of the two real forms of $sl(3,\mathbb{C})$. In fact, $sl(2,\mathbb{C})$ admits only a single non-compact real form since $su(1,1) \sim sl(2,\mathbb{R})$.

Since the algebra \eqref{sl3} is isomorphic to $sl(3,\mathbb{C})$, it is convenient to rewrite it in a more standard basis where one does not have to deal with the trace constraints on the generators. In particular, in the following we shall use the basis
\begin{subequations}\label{sl3can}
\begin{align}
& \left[\, L_i \comma L_j \,\right] \, = \, (i-j)\, L_{i+j} \, , \label{LL} \\[5pt]
& \left[\, L_i \comma W_m \,\right] \, = \, (2\,i-m)\, W_{i+m} \, , \label{LW} \\[5pt]
& \left[\, W_m \comma W_n \,\right] \, = \, \frac{\sigma}{3}\,(m-n)\,(\,2\,m^2+2\,n^2-mn-8\,)\, L_{m+n} \, ,  \label{WW}
\end{align}
\end{subequations}
where $-1 \leq i,j \leq 1$ and $-2 \leq m,n \leq 2$. It can be related to the previous one via the isomorphism
\be
J_0 \, = \, \frac{1}{2}\, (L_1 + L_{-1}) \, , \qquad J_1 \, = \, \frac{1}{2}\, (L_1 - L_{-1}) \, , \qquad J_2 \, = \, L_0 \, , \label{iso1}
\ee
that for the spin-3 generators reads
\begin{alignat}{3}
& T_{00} \, = \, \frac{1}{4} \left( W_2 + W_{-2} + 2\, W_0 \right) \, , \qquad\qquad & & T_{01} \, = \,  \frac{1}{4} \left( W_2 - W_{-2} \right) \, , \nn \\
& T_{11} \, = \, \frac{1}{4} \left( W_2 + W_{-2} - 2\, W_0 \right) \, , \qquad\qquad & & T_{02} \, = \,  \frac{1}{2} \left( W_1 + W_{-1} \right) \, , \nn \\
& T_{22} \, = \, W_0 \, , & & T_{12} \, = \,  \frac{1}{2} \left( W_1 - W_{-1} \right) \, . \label{iso2}
\end{alignat}
Notice that eq.\ \eqref{iso2} makes manifest the traceless condition
\be
- \, T_{00} \, + \, T_{11} \, + \, T_{22} \, = \, 0 \, . 
\ee
The $W_m$ generators thus provide a convenient parameterisation of the independent components of the $T_{ab}$ generators. The elimination of the trace constraints on the generators leads to important technical simplifications. In this respect, our choice represent an alternative to the spinorial notation adopted by Vasiliev \cite{vasiliev1} and Blencowe \cite{blencowe} to the same end.
  
In conclusion, one can describe the coupling of a spin-3 gauge field to AdS gravity in $D=\dim$ via the CS action \eqref{action} associated to the direct sum of two copies of the algebra \eqref{sl3}. Since $\sigma$ appears only in the commutator between two spin-3 generators it does not affect the linearised field equations. Therefore, it is possible to consider any direct sum of the two non-compact real forms of $sl(3,\mathbb{C})$. However, choosing the direct sum of the same real algebra entails a qualitative difference with respect to the choice of two different real forms. The distinction emerges when one performs the change of basis that induces the rewriting of the field equations in terms of the potentials 
\eqref{potentials}:
\begin{alignat}{3}
& M_a \, = \, J_a + \widetilde{J}_a \, , & & \qquad\qquad M_{ab} \, = \, T_{ab} \,+\, \widetilde{T}_{ab} \, , \nn \\[5pt]
& P_a \, = \, \frac{1}{l} \left(\, J_a - \widetilde{J}_a  \,\right) \, , & & \qquad\qquad P_{ab} \, = \, \frac{1}{l} \left(\, T_{ab} \,-\, \widetilde{T}_{ab} \,\right) \, , \label{PM}
\end{alignat}
where the tilde distinguishes the two copies of $sl(3,\mathbb{C})$. The commutators between the $P_a$ and the $M_a$ are not affected by the choice of $\sigma$ and $\tilde{\sigma}$ and one recovers the usual presentation of the $o(2,2)$ algebra:
\be
\left[\, M_a \comma M_b \,\right] \, = \, \e_{abc} M^c \, , \qquad  \left[\, M_a \comma P_b \,\right] \, = \, \e_{abc} P^c \, , \qquad \left[\, P_a \comma P_b \,\right] \, = \, \frac{1}{l^2}\, \e_{abc} M^c \, .
\ee
In a similar fashion, also the commutators mixing spin-2 and spin-3 generators are independent on the choice of $\sigma$ and $\tilde{\sigma}$:
\begin{alignat}{8}
& \left[\, M_a \comma M_{bc} \,\right] \, = \, \e^m{}_{a(b} M_{c)m} \, , \qquad\qquad & & \left[\, M_a \comma P_{bc} \,\right] \, = \, \e^m{}_{a(b} P_{c)m} \, , \\[5pt]
& \left[\, P_a \comma M_{bc} \,\right] \, = \, \e^m{}_{a(b} P_{c)m} \, , \qquad\qquad & & \left[\, P_a \comma P_{bc} \,\right] \, = \, \frac{1}{l^2}\, \e^m{}_{a(b} M_{c)m} \, .
\end{alignat}
On the other hand, the structure of the remaining commutators depends on $\sigma$ and $\tilde{\sigma}$ as
\begin{align}
& \left[\, M_{ab} \comma M_{cd} \,\right] \, = \, \12\, \big(\, \h_{a(c} \e_{d)bm} + \, \h_{b(c} \e_{d)am} \,\big)\, \big(\, (\sigma+\tilde{\sigma})\, M^m \, + \, l\,(\sigma-\tilde{\sigma})\, P^m \,\big) \, , \nn \\[5pt]
& \left[\, M_{ab} \comma P_{cd} \,\right] \, = \, \12\, \big(\, \h_{a(c} \e_{d)bm} + \, \h_{b(c} \e_{d)am} \,\big) \left(\, (\sigma+\tilde{\sigma})\, P^m \, + \, \frac{1}{l}\,(\sigma-\tilde{\sigma})\, M^m \,\right) \, , \nn \\[3pt]
& \left[\, P_{ab} \comma P_{cd} \,\right] \, = \, \12\, \big(\, \h_{a(c} \e_{d)bm} + \, \h_{b(c} \e_{d)am} \,\big) \left(\, \frac{1}{l^2}\, (\sigma+\tilde{\sigma})\, M^m \, + \, \frac{1}{l}\,(\sigma-\tilde{\sigma})\, P^m \,\right) \, . \label{comm_lim} 
\end{align}
By a suitable real rescaling of the generators the absolute values of $\sigma$ and $\tilde{\sigma}$ can be equated, but one can contract the full algebra with a $l \to \infty$ limit only if they have the same sign. A similar obstruction was recognised in \cite{townsend2} for the supergravity case, where one can consider two possible supergroups $OSp_\pm(\cN|2,\mathbb{R})$ depending on the overall sign of the anticommutator of the fermionic generators. In that case, to obtain a well defined Poincar{\'e} limit one has to choose two supergroups with opposite sign. At any rate, both choices could well be acceptable. For instance, in $D=4$ Vasiliev's theory for interacting higher-spin fields does not admit a flat-space limit \cite{review_vasiliev}.

We can now close this section by presenting the full non-linear action and the field equations in terms of the potentials \eqref{potentials}. This makes the interpretation of the CS theory as the coupling of a spin-3 field to gravity
more transparent. For simplicity we shall focus on $\sigma = \tilde{\sigma}$. 
In terms of the vielbein-like and of the spin-connection like fields the action \eqref{action} reads
\begin{align}
& S \, = \, \frac{1}{8 \p G}\int \bigg\{\, e^a \!\ww \left(\, d \o_a \, + \, \12\, \e_{abc}\, \o^b \!\ww \o^c \, - \, 2\, \sigma\, \e_{abc}\, \o^{bd} \!\ww \o^c{}_d \,\right) \label{action_explicit} \\
& -\, 2\, \sigma\, e^{ab} \!\ww \left(\, d \o_{ab} + \, \e_{cd(a|}\, \o^c \!\ww \o_{|b)}{}^d \,\right) + \, \frac{1}{6\,l^2}\, \e_{abc} \left(\, e^a \!\ww e^b \!\ww e^c - \, 12 \, \sigma\, e^a \!\ww e^{bd} \!\ww e^c{}_d \,\right) \!\bigg\} \, . \nn
\end{align}
Recall that for $\sigma > 0$ the gauge group is $SU(1,2)\times SU(1,2)$ while for $\sigma < 0$ it is $SL(3,\mathbb{R})\times SL(3,\mathbb{R})$. The equations of motion for the gravitational fields are
\begin{align}
& \mathscr{T}^a \equiv \, de^{\,a} \,+\, \e^{abc}\, \o_b \ww e_c \,-\, 4\,\sigma\, \e^{abc}\, e_{bd} \ww \o_c{}^d \, = \, 0 \, , \nn \\[5pt]
& \mathscr{R}^a \equiv \, d\o^a + \12\, \e^{abc} \left( \o_b \ww \o_c + \frac{e_b \ww e_c}{l^2} \right) - 2\,\sigma\, \e^{abc} \left(\, \o_{bd} \ww \o_c{}^d + \frac{e_{bd} \ww e_c{}^d}{l^2} \,\right) = \, 0 \, . \label{eq1}
\end{align}
Notice that the spin-3 fields provide a contribution to the torsion equation analogue to that appearing in $\cN=1$ supergravity \cite{townsend1,townsend2}. Indeed, the structure of the algebra \eqref{sl3} is very close to that of the $OSp_\pm(1|2,\mathbb{R})$ superalgebra, since the commutator of two spin-3 generators is proportional to a spin-2 generator. The field equations for the spin-3 fields are
\begin{align}
& \mathscr{T}^{ab} \equiv \, de^{\,ab} \,+\, \e^{cd(a|}\, \o_c \ww e_d{}^{|b)} \,+\, \e^{cd(a|} e_c \ww \o_d{}^{|b)} \, = \, 0 \, , \nn \\[5pt]
& \mathscr{R}^{ab} \equiv \, d\o^{ab} \,+\, \e^{cd(a|}\, \o_c \ww \o_d{}^{|b)} \,+\, \frac{1}{l^2}\, \e^{cd(a|} e_c \ww e_d{}^{|b)} \, = \, 0 \, . \label{eq2}
\end{align}
The coupling also deforms the gauge transformations. Besides the usual gauge transformations, the spin-2 fields acquire new gauge transformations proportional to the spin-3 gauge parameters $\x^{ab}$ and $\L^{ab}$:
\begin{align}
& \d\, e^{\,a} \, = \, - \, 4\,\sigma\, \e^{abc}\, \o_{bd}\, \x_{\,c}{}^d \,-\, 4\,\sigma\, \e^{abc}\, e_{bd}\, \L_c{}^d \, , \nn \\[5pt]
& \d\, \o^a \, = \,-\, 4\,\sigma\, \e^{abc}\, \o_{bd}\, \L_c{}^d \,-\, 4\, \frac{\sigma}{l^2}\, \e^{abc}\, e_{bd}\, \x_{\,c}{}^d \, . \label{gauge1} 
\end{align}
Similarly, the spin-3 fields also transforms under spin-2 gauge transformations. Their most general gauge transformations read
\begin{align}
& \d\, e^{\,ab} \, = \, d \x^{\,ab} \,+\, \e^{cd(a|}\, \o_c\, \x_{\,d}{}^{|b)} \,+\, \e^{cd(a|}\, e_c\, \L_d{}^{|b)} \,+\, \e^{cd(a} \o^{b)}{}_c\, \x_d \,+\, \e^{cd(a} e^{b)}{}_c\, \L_d \, , \nn \\[5pt]
& \d\, \o^{ab} \, = \, d \L^{\,ab} \,+\, \e^{cd(a|}\, \o_c\, \L_{d}{}^{|b)} \,+\, \frac{1}{l^2}\, \e^{cd(a|}\, e_c\, \x_{\,d}{}^{|b)} \,+\, \e^{cd(a} \o^{b)}{}_c\, \L_d \,+\, \frac{1}{l^2}\, \e^{cd(a} e^{b)}{}_c\, \x_d \, . \label{gauge2}
\end{align}
%

\section{Review of Chern-Simons theory with boundary}\label{sec:boundary}

This section is a review of well known facts on CS theory in $D=3$, 
which we include for completeness of the presentation and to fix the notation. We largely follow the expositions in \cite{banados_CS,review_banados,carlip}.

Consider a generic Chern-Simons theory with gauge group $G$ (generated by the Lie algebra $\mathfrak{g}$) on a space $\cM = \mathbb{R}\times \Sigma$, where $\Sigma$ is a two-dimensional manifold
with boundary $\partial \Sigma \cong S^{1}$. On manifolds with boundary the CS action \eqref{chern-simons} in general is neither differentiable nor gauge invariant. In fact, when one varies it, one obtains the boundary contribution
\begin{equation}\label{CSvariation}
\delta S_{CS}\Big|_{\text{boundary}} = 
- \,\frac{k}{4\pi} \int_{\mathbb{R}\times S^{1}} \tr (A\wedge \delta A) \ .
\end{equation}
In order to have a well defined action principle one either has to add
boundary terms to the action \eqref{chern-simons} or to impose
suitable boundary conditions on fields. As we shall see, the choice of
boundary conditions will play a crucial role in our higher-spin setup.
The presence of a timelike boundary affects the phase space of the
theory. In general it becomes infinite dimensional, and there are 
infinitely many global charges satisfying an algebra that depends on the
choice of boundary conditions. To find the algebra of global charges
we follow the method of Regge and Teitelboim \cite{regge_teitelboim} as
applied to the CS theory in \cite{banados_CS,review_banados}.

The first step is to rewrite the CS action using a $(2+1)$-decomposition of the gauge field,
\begin{equation}
A \,=\, A_{0}\,dt \,+\, A_{i}\,dx^{i} \ .
\end{equation}
The action then reads
\begin{equation}
S_{CS} = \frac{k}{4\pi} \int_\cM dt\wedge dx^{i}\wedge dx^{j}\, \tr
\big(A_{0}F_{ij} - A_{i}\dot{A}_{j} \big) + \frac{k}{4\pi} 
\int_{\mathbb{R}\times S^1} dt\wedge dx^{i}\, \tr
\big(A_{0}A_{i} \big) \ . 
\end{equation}
This action has $2N$ dynamical fields $A_{i}$ (where $N$ is the
dimension of the gauge group $G$) and $N$ Lagrange multipliers
$A_{0}$. The field equations of the Lagrange multipliers provide first class constraints which generate gauge transformations. The equal-time Poisson bracket of two differentiable phase-space functionals $F[A_i]$ and $H[A_i]$ is defined by
\begin{equation} \label{poisson_original}
\{F,H\} \,=\, \frac{2\pi}{k} \int_\Sigma dx^i \ww dx^j \,
\tr\left( \frac{\delta F}{\delta A_i(x)} \, \frac{\delta
H}{\delta A_j(x)} \right) \, .
\end{equation}
One defines the smeared generators of gauge transformations 
\be \label{smearedgen0}
G(\Lambda) \, = \, \frac{k}{4\pi} \int_\S dx^i \ww dx^j \,\tr(\Lambda\, F_{ij}) \, 
+ \, Q(\Lambda) \, ,
\ee
where $Q(\Lambda)$ is a boundary term whose role is to 
cancel the surface term that arises if one writes the variation of the first term in 
eq.\ \eqref{smearedgen0} in terms of $\delta A_i$ rather than its derivatives
\cite{BCT}.
 
True (proper) gauge transformations are those for which the surface term vanishes.
If the gauge parameter $\Lambda$  is independent of the fields, the boundary term takes the form
\be \label{smearedgen}
Q(\Lambda) \, = \, - \, \frac{k}{2\pi} \int_{\pr\S} dx^i \,\tr(\Lambda\, A_i) \, . 
\ee
This leads to the Poisson algebra
\be \label{poissongen}
\left\{ G(\Lambda) , G(\Gamma) \right\} \, = \, G([\Lambda,\Gamma]) \, 
+ \, \frac{k}{2\pi} \int_{\pr\S} dx^i\, \tr(\Lambda\,\pr_i\Gamma)  \, ,
\ee
where the central extension crucially rests on the presence of the surface term 
$Q(\Lambda)$ in the definition of the smeared generator. Notice that the boundary contribution $Q(\Lambda)$ does not vanish when the 
constraints $F_{ij}=0$ are imposed. 
The transformations generated by a $G(\Lambda)$ with $\Lambda$
such that $Q(\Lambda)$ is non-zero, 
are not true gauge transformations, but rather global symmetries which 
transform physically inequivalent configurations into each other. 
This is also the origin of the infinitely many boundary degrees of freedom. 

After gauge fixing and solving the constraints,  
the $Q(\Lambda)$ define the global charges of the CS theory.
They generate global symmetries by acting on a generic phase-space functional $F$ as
\be \label{chargegen}
\d_{\L} F \, = \, \left\{ Q(\L) , F \right\} \, ,
\ee
and they satisfy the same algebra as the $G(\Lambda)$, i.e. 
\be \label{chargepoisson}
\left\{ Q(\L) , Q(\G) \right\} \, = \, Q([\L,\G]) \, 
+ \, \frac{k}{2\pi} \int_{\pr\S} dx^i\, \tr(\L\,\pr_i\G)  \, ,
\ee 
but now the brackets are Dirac brackets on the reduced phase space. 

We now present a set of boundary conditions which ensure
the differentiability of the CS action and a gauge-fixing procedure
that will play a crucial role in the following. To select the boundary
conditions it is convenient to introduce light cone coordinates
$x^{\pm} = \frac{t}{l}\pm \theta$, where $t$ parameterises the time
direction while $\theta$ parameterises the circle at the
boundary. Eq.\ \eqref{CSvariation} becomes 
\begin{equation} \label{boundvarlight}
\delta S_{CS}\big|_{\text{boundary}} = -\,\frac{k}{4\pi} \int_{\mathbb{R}\times S^1} dx^{+}dx^{-}\,\tr \big(A_{+}\delta A_{-} -A_{-}\delta A_{+} \big)\ .
\end{equation}
This vanishes if we impose
\begin{equation} \label{boundary1}
A_{-} = 0 \quad \text{at the boundary.}
\end{equation}
We shall later see that this choice can also be motivated from the
gravity description: for instance, all black hole solutions and whatever is generated from them by the action of asymptotic Killing vectors/tensors satisfy this condition. 

Let us now assume that the constant time slices $\Sigma$ have the topology of a disc, which we parameterise by a radial coordinate $\r$ and the previous angle variable $\theta$. To fix the gauge, we choose a function $b (\r)$ with values in the group $G$ and we set
\begin{equation}\label{gaugechoice}
A_{\r} \,=\, b^{-1} (\r)\,\partial_{\r}\, b (\r) \ .
\end{equation}
This choice is always possible. Assume we start with a gauge field
$A'$, and we want to perform a gauge transformation $U$ to bring it to the form~\eqref{gaugechoice},
\begin{equation}\label{realisationofgaugechoice}
U^{-1}A'_{\r}\,U \,+\, U^{-1}\partial_{\r}U \,=\, b^{-1}\partial_{\r}b \ .
\end{equation}
We write $U=U'b$, and from~\eqref{realisationofgaugechoice} we obtain
\begin{equation}
\partial_{\r}U' \,=\, -\, A'_{\r}\,U' \ ,
\end{equation}
which can be solved by a path-ordered exponential,
\begin{equation}
U \,=\, \cP e^{-\int^{\r} A'_{\r}d\r'} U_{0} b \ .
\end{equation}
$U_{0}$ is a constant of integration (independent of $\r$) and is
chosen such that $U=1$ at the boundary. This leaves the boundary condition
\eqref{boundary1} untouched. 
Therefore, $U$ is an allowed
gauge transformation also in the theory with boundary, and the gauge
choice~\eqref{gaugechoice} is always possible.

From the constraint $F_{\r\theta}=0$ we then find
\begin{equation}
\partial_{\r}A_{\theta} \,+\, [A_{\r},A_{\theta}] \,=\, 0 \ ,
\end{equation}
which is solved by 
\begin{equation} \label{solgaugefixed}
A_{\theta} (t,\r,\theta) \,=\, b^{-1} (\r)\, a (t,\theta)\, b (\r)\ .
\end{equation}
The $\r$-dependence of the Lagrange multiplier $A_{0}$ is determined by the equation of
motion for the gauge-fixed (and therefore constant) $A_{\r}$,
\begin{equation}
\partial_{\r}A_{0} \,+\, [A_{\r},A_{0}] \,=\, 0 \ .
\end{equation}
The boundary condition $A_{-}=\frac{1}{2}
(A_{0}-A_{\theta})=0$ then forces $A_{0}$ to coincide with
$A_{\theta}$ everywhere; in other words on shell
\begin{equation} \label{boundary1_2}
A_{-} \,\equiv\, 0 
\end{equation}
everywhere, not only at the boundary. Therefore, the phase space is parameterised by $a(t,\theta)$. The gauge choice \eqref{gaugechoice} is preserved by gauge transformations whose parameters are of the form
\be \label{gaugeparglobal}
\L(t,\r,\theta) \, = \, b^{-1} (\r)\, \l (t,\theta)\, b (\r)\quad\Rightarrow\quad
\delta A_\rho=0\ ,
\ee 
but the boundary condition \eqref{boundary1} implies that they generate
global symmetries. In fact, an arbitrary time dependence for $\l(t,\theta)$ is not compatible with \eqref{boundary1}:
\be
\d A_- \,=\, 0 \quad \Rightarrow \quad \pr_- \l \,=\, 0 \, . 
\ee
This is also the condition on the gauge parameter under which the 
CS action is gauge invariant.
Similarly, the on-shell condition on $a(t,\theta)$ is that it depends only on $x^+$. This confirms the absence of an arbitrary time dependence and therefore proves that \eqref{gaugechoice} is not only admissible, but also completely fixes the gauge freedom. 

The  Poisson structure on the reduced phase space can be obtained by
inserting in eq.\ \eqref{chargepoisson} the gauge-fixed expression for the charges \eqref{smearedgen},
\be \label{charges}
Q(\l) \, = \, - \, \frac{k}{2\pi} \int d\th\ \tr \left(\l(\th)\,a(\theta)\right) \, . 
\ee
Expanding $a(\theta)$ in a basis $\{\cT_{A}\}$ of the Lie
algebra, $a= a^{A}\cT_{A}$, one obtains the affine Lie algebra
\be
\left\{ a^A(\theta) , a^B(\theta^\pe) \right\} \,=\, - \,\frac{2\pi}{k} 
\Big(\, \d(\theta-\theta^\pe)\, f^{AB}{}_C\, a^C(\theta) \, 
- \, \d^\pe(\theta-\theta^\pe)\, \g^{AB}  \,\Big) \, ,
\ee
where $f_{AB}{}^C$ are the structure constant of $\mathfrak{g}$ and 
$\gamma^{AB}$ is the inverse of the Killing metric $\gamma_{AB}$. The Killing metric and its inverse are used to lower and raise the indices of the structure constants. Decomposing $a(\theta)$ into Fourier modes,
\begin{equation}
a^A (\theta) \,=\, \frac{1}{k}\sum_{p\in \mathbb{Z}}\, a^A_{p}\, e^{-ip \theta}\ ,
\end{equation}
leads to 
\begin{equation}\label{affinealgebra}
\{a^{A}_{p},a^{B}_{q} \} \,=\, -\, f^{AB}{}_{C}\,a^{C}_{p+q} \,+\,
ipk\,\gamma^{AB}\,\delta_{p+q,0}\ .
\end{equation}

\section{Asymptotic symmetries}\label{sec:symmetries}

In this section we analyse the asymptotic symmetries of gravity
coupled to a spin-3 field in backgrounds which are asymptotically AdS.
We work in the frame-like formulation as a CS theory. First we
motivate the boundary conditions we have to impose in the CS theory by
reconsidering those of pure gravity, and then we define when a
configuration is \textit{asymptotically AdS} in our higher-spin
context. This leads to the identification of the Poisson structure on
this solution space. The Poisson algebra structure we find is that of
a classical centrally extended $\cW_3 \otimes \cW_3$ algebra. Finally,
we discuss the implications of our boundary conditions in the
metric-like formulation.

\subsection{Boundary conditions}\label{sec:bound_cond}

As anticipated in Section \ref{sec:boundary}, the boundary condition
\eqref{boundary1} emerges naturally when describing asymptotically
Anti-de Sitter solutions of the field equations. On the other hand, it is not sufficient to fully characterise this class of solutions, and it has to be supplemented by further boundary conditions. We shall now confirm the role of \eqref{boundary1} and identify the extra requirements by a close scrutiny at the properties of the asymptotically AdS solutions of Einstein gravity. In fact, any solution of the field equations of pure gravity can be embedded in any CS higher-spin extension. It simply corresponds to a solution where all higher-spin fields vanish.\footnote{The embedding of BTZ solutions into three-dimensional higher-spin gauge theories was already considered in \cite{BTZ_HS} as a playground for the study of exact solutions in this context.} Therefore, these pure-gravity backgrounds provide a subset of the space of asymptotically AdS solutions we are going to characterise.

In \cite{review_banados} it was pointed out that the metric
\be \label{sol_einstein}
ds^2 = \, l^2 \left\{\, d\rho^2 - \frac{8\pi G}{l} \left(\, \cL\, (dx^+)^2 + \widetilde{\cL}\, (dx^-)^2  \,\right) - \left(\, e^{2\rho} + \frac{64\pi^2G^2}{l^2}\,\cL\,\widetilde{\cL}\, e^{-2\rho}  \,\right) dx^+ dx^- \,\right\}
\ee
is an exact solution of Einstein's equations for any $\cL = \cL(x^+)$ and $\widetilde{\cL} = \widetilde{\cL}(x^-)$. Moreover, it parameterises the whole space 
of asymptotically ($\rho\to\infty$)
AdS solutions with a flat boundary metric.\footnote{This is, in fact, 
a special case of the general result \cite{skenderis} that in three dimensions
the Fefferman-Graham expansion \cite{FG} terminates after the third term.} 
For instance, if we set the two functions 
$\cL$ and $\widetilde{\cL}$ to the constant values
\be \label{BTZ_L}
\cL_0 \, = \, -\, \frac{1}{4\pi} \, (Ml-J) \, , \qquad\qquad \widetilde{\cL}_0 \, = \, - \, \frac{1}{4\pi} \, (Ml+J) \, ,
\ee
eq.\ \eqref{sol_einstein} is the BTZ solution with mass $M>0$ and angular momentum $|J|\leq Ml$. For $8GM = -1$ and $J=0$ it is the $AdS_3$ solution. The change of coordinates casting the BTZ metric \eqref{sol_einstein} in the usual form in terms of lapse and shift functions can be found in \cite{review_banados}. 

Introducing $b (\r) = e^{\r L_{0}}$ the metric \eqref{sol_einstein} can be described by the connections
\begin{align} \label{BTZ}
& A \, = \, b^{-1} \left(\, L_{1} \, + \, \frac{2\pi}{k}\, \cL(x^+) \, L_{-1}  \,\right) b \, dx^+ \, + \, b^{-1} \,\pr_{\r}\, b\, d \r \, , \nn \\
& \widetilde{A} \, = \, - \, b \left(\, \frac{2\pi}{k}\, \widetilde{\cL}(x^-) \, L_{1} \, + \ L_{-1} \,\right) b^{-1} \,
  dx^- \, + \,  b \,\pr_{\r}\, b^{-1}\, d \r \, ,
\end{align}
that are related to the dreibein and the spin connection through eqs.\
\eqref{gravity_potentials} and \eqref{forms}. Notice that we exploited the relation \eqref{level} between the level of the CS action and the AdS radius. We also resorted to the basis \eqref{sl3can} that will prove particularly convenient when we shall extend the discussion to the whole $SL(3) \times SL(3)$ CS theory.

The connections \eqref{BTZ}, which translate the Brown-Henneaux boundary conditions into the frame formalism,  were introduced in  
\cite{henneaux_vandriel} and it was pointed out in
\cite{review_banados} that they provide an exact solution of the
Einstein equations. Solutions parameterised by different functions
$\cL$ and $\widetilde{\cL}$ cannot be related by proper gauge transformations, and thus they are physically inequivalent. 
This characterisation of the space of asymptotically AdS solutions was extended 
to supergravity theories with one or several 
spin-3/2 fields in \cite{sugra_asymptotics}. Our generalisation consists in including a spin-3 field which 
is coupled to gravity. Further possible generalisation with several fields 
with integer and/or half-integer spins $>2$ based on higher 
rank groups and supergroups should be straightforward, but we will not 
explicitly consider them here; see however the discussion in Section 
\ref{sec:generic_group}. 

Notice that the connection $A$ of eq.\ \eqref{BTZ} satisfies the gauge choice \eqref{gaugechoice} and the condition \eqref{boundary1_2} in the whole space and 
$\widetilde{A}$ satisfies analogous conditions. We shall now show that these properties continue to hold on the wider space of solutions which is obtained by acting on a generic pure-gravity background of the form \eqref{BTZ} with the isometries of the $AdS_3$ solution. As such, they can be considered as crucial ingredients in the characterisation of generic asymptotically AdS solutions even in the full $SL(3)\times SL(3)$ CS theory.

To prove this statement, notice that in the AdS case, for which
$\frac{2\pi}{k}\cL=\frac{2\pi}{k}\widetilde\cL=\frac{1}{4}$,
eq.\ \eqref{BTZ} can be written as
\be
A \, = \, g^{-1} d\, g \, , \qquad\qquad \widetilde{A} \, = \, \tilde{g}^{-1} d\, \tilde{g} \, ,
\ee
with
\be
g \, = \, e^{\frac{x^+}{2} \left(\,L_1+L_{-1}\,\right)}\, b(\r) \, , \qquad\qquad \tilde{g} \, = \, e^{-\frac{x^-}{2} \left(\, L_1+L_{-1}\,\right)}\, b^{-1}(\r) \, .
\ee
With this rewriting we can present the isometries of 
$AdS_3$ in a very compact form and we can relate them to the $AdS_3$ Killing vectors and tensors. For instance, if we denote collectively the $SL(3)$ generators by $\cT_A$, gauge transformations generated by the parameters
\be \label{iso_par}
\x_A \, = \, g^{-1}\, \cT_A \, g \, = \, \left(\, g^{-1}\, \cT_A \, g \,\right)^B\, \cT_B
\ee
leave the $AdS_3$ connection $A$ invariant and similarly for $\widetilde A$.
If we contract the gauge parameters with the inverse of the AdS dreibein, e.g.
\be \label{killing}
v^{\m}_a \, = \, \bar{e}^{\,\m}{}_b\, \left(\, g^{-1}\, J_a \, g \,\right)^b \, ,  \qquad\qquad
k^{\m\n}_{ab} \, = \, \bar{e}^{\,\m}{}_c\, \bar{e}^{\,\n}{}_d\, \left(\, g^{-1}\, T_{ab} \, g \,\right)^{cd} \, ,
\ee
the complete set of isometries gives rise to the 6 Killing vectors of $AdS_3$ and to its 10 traceless Killing tensors. The latter can be interpreted as the generators of spin-3 gauge transformations \eqref{gauge_metric} which leave the 
AdS background invariant.
If one acts repeatedly with the gauge transformations generated by \eqref{iso_par} (or their analogues involving $\tilde{g}$) on a generic solution \eqref{BTZ} one obtains expressions of the form
\begin{alignat}{3}
& A_+ \, = \, b^{-1}(\r) \,a(x^+)\, b(\r) \, , \qquad\qquad & & \widetilde{A}_+ \, = \, 0 \, , \nn \\
& A_- \, = \, 0 \, , \qquad\qquad & & \widetilde{A}_- \, = \, b(\r) \,\tilde{a}(x^-)\, b^{-1}(\r) \, , \nn \\
& A_\r \, = \, b^{-1}(\r) \,\pr_{\r}\, b(\r) \, , \qquad\qquad & & \widetilde{A}_\r \, = \,  b(\r) \,\pr_{\r}\, b^{-1}(\r) \, , \label{affine_sol}
\end{alignat}
with $b=e^{\rho L_0}$. 
$a(x^+)$ and $\tilde{a}(x^-)$ are Lie-algebra valued functions 
which take values in the whole $sl(3)$. For instance
\be
a(x^+) \, = \, \sum_{i=-1}^{1}\ell^{\,i} (x^+) \, L_{i} + \sum_{m=-2}^{2}
w^{\,m} (x^+) \, W_{m} \, .
\ee
As anticipated, the flat connections \eqref{affine_sol} still satisfy the conditions 
\eqref{boundary1_2} and \eqref{gaugechoice}. 

Nevertheless, eq.\ \eqref{affine_sol} does not yet provide a
satisfactory parameterisation of the space of asymptotically AdS
solutions. In fact, the discussion of Section \ref{sec:boundary} makes
it clear that the asymptotic symmetries of the solution space
\eqref{affine_sol} are described by a Kac-Moody algebra. On the other
hand, the most natural way to define asymptotically AdS solutions in
any extension of Einstein gravity is to keep the asymptotic conformal
symmetry of pure gravity and perhaps to extend it. Moreover, the
procedure leading to \eqref{affine_sol} destroys the parameterisation
\eqref{BTZ} even in the pure gravity sector. This strongly suggests
that eq.\ \eqref{affine_sol} should be supplemented by additional boundary conditions analogous to those introduced in \cite{henneaux_vandriel} for pure gravity. 

In the gravity sector this can be understood by noticing that
diffeomorphisms actually constitute a particular class of gauge
transformations: those with gauge parameters linear in the fields
\cite{witten}. This enables one to recover the conformal
asymptotic symmetry group of pure gravity directly in the CS
formulation \cite{banados_CS} and this still holds for higher-spin
gauge transformations. For instance, by inverting eq.\ \eqref{par_lin}
(for $s=3$) one realises that, at the linearised level, the
metric-like gauge transformations \eqref{gauge_metric} are associated
to CS gauge transformations whose parameters are quadratic in the
inverse dreibein. Unfortunately, when considering non-trivial
spin-3 backgrounds the identification between metric-like gauge
transformations -- which are the spin-3 analogue of the gravity
diffeomorphism -- and CS gauge transformations becomes definitely more
complicated.

Rather than giving a detailed characterisation of this relation in the spirit of 
\cite{banados_CS}, we can identify the additional boundary conditions simply by requiring that the resulting asymptotic symmetry group contains the conformal group. This can be achieved by looking at the Anti-de Sitter solution itself,
\begin{align} \label{AdSconnection}
& A_{AdS} \, = \, b^{-1} \left(\, L_{1} \, + \, \frac{1}{4} \, L_{-1}  \,\right) b \, dx^+ \, + \, b^{-1} \,\pr_{\r}\, b\, d \r \, , \nn \\
& \widetilde{A}_{AdS} \, = \, - \ b \left(\, \frac{1}{4} \, L_{1} \, + \ L_{-1} \,\right) b^{-1} \,
  dx^- \, + \,  b \,\pr_{\r}\, b^{-1}\, d \r \, ,
\end{align}
and by constraining the deviation of a generic solution of the field equations from its boundary value. We thus call a solution \textit{asymptotically Anti-de Sitter} if it satisfies~\eqref{gaugechoice} and~\eqref{boundary1_2} and its difference to the AdS-solution is finite at the boundary,
\begin{equation}\label{AdScondition}
\left(\, A \, - \, A_{\text{AdS}} \,\right)\Big|_{\text{boundary}} = \, \mathcal{O} (1)\ ,
\end{equation}
with a similar condition for $\widetilde{A}$.
In the following we will mainly deal with $A$, the results for 
$\widetilde A$ follow by analogy.

In the next section we shall prove that any background which satisfies 
the rather natural additional boundary condition \eqref{AdScondition}  has an  asymptotic extended conformal symmetry.
This result is the main motivation for choosing the boundary conditions \eqref{boundary1} and \eqref{AdScondition} in the gauge \eqref{gaugechoice}. 
Notice that they are satisfied by the pure-gravity solutions \eqref{BTZ} and by their supergravity extensions \cite{sugra_asymptotics}. In Section \ref{sec:metric-like} we shall express them as fall-off conditions for the metric-like fields, thus showing their analogies with the standard Brown-Henneaux boundary conditions.

\subsection{Asymptotic symmetry algebra}

In Section \ref{sec:boundary} we have seen that, after the gauge fixing \eqref{gaugechoice}, for any CS theory 
the space of solutions satisfying the boundary condition \eqref{boundary1} is parameterised by $a (\theta)$. 
For $SL(3)$ we expand $a(\theta)$ in the $\{L_{i},W_{m}\}$ basis,
\begin{equation}
a (\theta) \, = \sum_{i=-1}^{1}\ell^{\,i} (\theta) L_{i} + \sum_{m=-2}^{2}
w^{\,m} (\theta) W_{m} \ .
\end{equation}
The additional boundary condition \eqref{AdScondition} then translates into the following conditions on the components $\ell^{i}$ and $w^{m}$:
\begin{equation}\label{AdSconditiononcomponents}
\ell^{1} = 1 \ , \qquad w^{1}=w^{2}= 0 \ .
\end{equation}
These are first-class constraints with respect to the Poisson
structure \eqref{affinealgebra} (given explicitly in
Appendix~\ref{sec:formulae}), and therefore they generate gauge
transformations. We can use them to set
\begin{equation}\label{AdSconditiongaugefixing}
\ell^{0} = 0\ , \qquad w^{0}=w^{-1} =0 \ .
\end{equation}
This completely fixes the gauge freedom \cite{WZW-Toda}, and the set of
constraints~\eqref{AdSconditiononcomponents}
and~\eqref{AdSconditiongaugefixing} is now second class. The degrees
of freedom that remain are the components $\ell^{-1}$ and
$w^{-2}$. Different choices for them distinguish physically inequivalent solutions.

We can now choose a convenient normalisation for these components and cast a generic asymptotically Anti-de Sitter solution of the field equations in the form
\be \label{sol_gen}
a(\theta) \, = \, L_1 \, + \, \frac{2\pi}{k}\, \cL(\theta)\, L_{-1} \, + \, \frac{\pi}{2k\sigma}\, \cW(\theta)\, W_{-2} \, .
\ee
This extends eq.\ \eqref{BTZ} by the addition of the $\cW(\theta)$ term. 
Below eq.\ \eqref{gaugeparglobal} we have seen that the global symmetries of the space of solutions parameterised by $a(\theta)$ are described by global gauge transformations. We expand their parameters as
\be
\l(\theta) \, = \, \sum_{i=-1}^{1}\ve^{\,i} (\theta) L_{i} + \sum_{m=-2}^{2}
\c^{\,m} (\theta) W_{m}
\ee
and identify those that leave the structure 
of \eqref{sol_gen} invariant. They are characterised by relations between the $\ve^i$ and $\c^m$ which are conveniently expressed in terms of $\ve \equiv \ve^{\,1}$ and $\c \equiv \c^{\,2}$. Denoting $\theta$-derivatives by primes, one obtains
\begin{align}
& \ve^{\,0} \, = \, - \, \ve^\pe \, , \nn \\
& \ve^{\,-1} \, = \, \12\, \ve^{\pe\pe} \, + \, \frac{2\p}{k}\, \ve\,\cL \, + \, \frac{4\p}{k}\, \c\,\cW \, , 
\end{align}
and
\begin{align}
& \c^{\,1} \, = \, - \, \c^{\,\pe} \, , \nn \\
& \c^{\,0} \, = \, \12\, \c^{\,\pe\pe} \, + \, \frac{4\p}{k}\, \c\,\cL \, , \nn \\
& \c^{\,-1} \, = \, -\, \frac{1}{6}\, \c^{\,\pe\pe\pe} \, - \, \frac{10\p}{3k}\, \c^{\,\pe}\cL \, - \, \frac{4\p}{3k}\, \c\,\cL^\pe \, , \nn \\
& \c^{\,-2} \, = \, \frac{1}{24}\, \c^{\,\pe\pe\pe\pe} \, + \, \frac{4\pi}{3k}\, \c^{\,\pe\pe}\cL \, + \, \frac{7\pi}{6k}\, \c^{\,\pe}\cL^\pe \, + \, \frac{\p}{3k}\, \c\,\cL^{\pe\pe} \, + \, \frac{4\pi^2}{k^2}\, \c\,\cL^2 \, + \, \frac{\p}{2k\sigma} \, \ve\,\cW \, .
\end{align}
Under these transformations the functions $\cL(\theta)$ and $\cW(\theta)$ vary as
\begin{subequations}\label{var_ea}
\begin{align}
& \d_\ve\, \cL \, = \, \ve\, \cL^\pe \, + \, 2\ \ve^\pe \cL \, + \, \frac{k}{4\pi}\, \ve^{\pe\pe\pe} \, , \label{d_e_L} \\
& \d_\ve\, \cW \, = \, \ve\, \cW^{\,\pe} \, + \, 3\ \ve^\pe\, \cW \, , \label{d_e_W}
\end{align}
\end{subequations}
and
\begin{subequations}\label{var_ca}
\begin{align}
& \d_\c\, \cL \, = \, 2\, \c\,\cW^{\,\pe} \, + \, 3\, \c^{\,\pe}\, \cW \, , \label{d_c_L} \\
& \d_\c\, \cW  \, = \, \frac{\sigma}{3} \, \bigg(\, 2\,\c\,\cL^{\pe\pe\pe} \,+\, 9\,\c^{\,\pe}\cL^{\pe\pe} \,+\, 15\,\c^{\,\pe\pe}\cL^\pe \,+\, 10\,\c^{\,\pe\pe\pe}\cL \,+\, \frac{k}{4\pi}\, \c^{(5)} \nn \\
& \phantom{\d_\c\, \cW  \, = \,} + \frac{64\pi}{k} \left(\, \c\,\cL\cL^\pe \,+\, \c^{\,\pe}\cL^2 \,\right) \bigg) \, . \label{d_c_W}
\end{align}
\end{subequations}
Notice that eq.\ \eqref{d_e_W} manifests that $\cW$ is a primary field of conformal weight 3 with respect to $\cL$, that plays the role of  energy momentum tensor for the boundary theory. This is actually one of the main advantages of the gauge fixing that leads to \eqref{AdSconditiongaugefixing}. With the convention \eqref{killingmetric} for the Killing metric (which is consistent with 
\eqref{convention_killing}) the charges \eqref{charges} which generate the transformations \eqref{var_ea} and \eqref{var_ca} read
\begin{align}
& Q(\lambda) \, = \int d\th\, \ve(\th)\, \cL(\th) \, +\int d\th\, \c(\th)\, \cW(\th) \, .
\end{align}
They generate global symmetries via eq.\ \eqref{chargegen} which
allows us to identify the Poisson structure on the phase space of
asymptotically Anti-de Sitter solutions. Eqs.\ \eqref{var_ea} and
\eqref{var_ca} lead to
\begin{subequations}\label{W3}
\begin{align}
& \left\{ \cL(\th) , \cL(\th^\pe) \right\} \, = \, - \, \left(\, \d(\th-\th^\pe) \cL^\pe(\th) \,+\, 2\,\d^\pe(\th-\th^\pe)\cL(\th) \,\right) \, - \, \frac{k}{4\pi}\, \d^{\pe\pe\pe}(\th-\th^\pe)\,\, , \\[5pt]
& \left\{ \cL(\th) , \cW(\th^\pe) \right\} \, = \, - \, \left(\, 2\,\d(\th-\th^\pe) \cW^\pe(\th) \,+\, 3\,\d^\pe(\th-\th^\pe)\cW(\th) \,\right)\, , \\[10pt]
& \left\{ \cW(\th) , \cW(\th^\pe) \right\} \, = \, \nn \\
& - \, \frac{\sigma}{3} \,\bigg(\, 2\,\d(\th-\th^\pe)\cL^{\pe\pe\pe}(\theta) + 9\,\d^\pe(\th-\th^\pe)\cL^{\pe\pe}(\th) + 15\,\d^{\pe\pe}(\th-\th^\pe)\cL^\pe(\th) + 10\,\d^{\pe\pe\pe}(\th-\th^\pe)\cL(\th) \quad \nn \\ 
& + \frac{k}{4\pi}\, \d^{(5)}(\th-\th^\pe)\, + \frac{64\pi}{k} \left(\, \d(\th-\th^\pe)\cL(\th)\cL^\pe(\th) + \d^\pe(\th-\th^\pe)\cL^2(\th) \,\right) \,\bigg) \, .
\end{align}
\end{subequations}
This is the classical $\cW_3$-algebra (see e.g. \cite{mathieu}) with central charge
\be
\label{centralcharge}
c \,=\, 6\,k \,=\, \frac{3\,l}{2\,G} \, ,
\ee
which is the same as for pure gravity \cite{BH}. 

An alternative way to present the $\cW_3$-algebra is in terms of the Fourier modes
of $\cL$ and $\cW$ which are defined as
\be
\cL(\theta) \, = \, - \, \frac{1}{2\p} \sum_{p \in \mathbb{Z}} \, \cL_p \, e^{-ip\theta} \, , 
\qquad\qquad \cW(\theta) \, = \, \frac{1}{2\p} \sum_{p \in \mathbb{Z}} \, 
\cW_p \, e^{-ip\theta} \, .
\ee
If we shift the $\cL$ zero mode according to 
\be
\cL_p \, \to \, \cL_p \, - \, \frac{k}{4}\, \d_{p,0}
\ee
and use  $c = 6k$, we obtain\footnote{This differs from eq.\ (19) of \cite{mathieu}  by a rescaling of the $\cW_n$ by a factor of  $\sqrt{10}$.}
\begin{subequations}\label{W3fourier1}
\begin{align}
& i \left\{\, \cL_p \comma \cL_q \,\right\} \, = \, (p-q)\, \cL_{p+q} \, + \, \frac{c}{12} \, (p^3-p) \, \d_{p+q,0} \, , \label{LL_fourier} \\
& i \left\{\, \cL_p \comma \cW_q \,\right\} \, = \, (2p-q)\, \cW_{p+q} \, , \label{LW_fourier} \\
& i \left\{\, \cW_p \comma \cW_q \,\right\} \, = \, - \, \frac{\sigma}{3} \,\bigg[\, (p-q)(2p^2+2q^2-pq-8)\, \cL_{p+q} \, + \, \frac{96}{c}\, (p-q)\,\L_{p+q} \phantom{(4.31b)} \nn \\
& \phantom{i \left\{\, \cW_p \comma \cW_q \,\right\} \, = \, - \, \frac{\sigma}{3} \,\bigg[\,} + \, \frac{c}{12}\, p(p^2-1)(p^2-4)\, \d_{p+q,0} \,\bigg] \, , \label{WW_fourier}
\end{align}
\end{subequations}
where we have defined
\begin{equation}
\Lambda_{p} \,\equiv\, \sum_{q\in\mathbb{Z}}\, \cL_{p+q} \cL_{-q}\ .
\end{equation}
The same algebra is obtained for each of the two  $SL(3)$ CS theories
which comprise the action \eqref{action}.
Therefore, the asymptotic symmetry of a spin-3 field coupled to gravity 
which is asymptotically AdS generate the  $\cW_3 \otimes \cW_3$ algebra.  

The approach we followed in deriving \eqref{W3fourier1} is the one used e.g. in \cite{sugra_asymptotics}. 
We now present, following \cite{review_banados}, an alternative derivation
by explicitly computing the Dirac brackets of the generators of the algebra.
If we collectively denote the second-class constrains 
\eqref{AdSconditiononcomponents} and \eqref{AdSconditiongaugefixing} 
by $\{\chi_{\alpha}\approx 0 \}$, we need to compute, using 
\eqref{affinealgebra}, the (non-degenerate) 
matrix $C_{\alpha\beta}=\{\chi_{\alpha},\chi_{\beta} \}$. 
The Dirac bracket between two phase-space functions $f,\,g$ on the constraint surface 
is
\begin{equation}
\{ f,g\}_{*} = \{f,g \} - \{f,\chi_{\alpha } \}\big(C^{-1}
\big)^{\alpha \beta} \{\chi_{\beta},g \} \ .
\end{equation}
We work directly with the Fourier modes of $\ell^m(\theta)$ and $w^n(\theta)$, defined as
\begin{equation} \label{modesW}
\ell^{m} (\theta) = \frac{1}{k} \sum_{p\in \mathbb{Z}}
\ell^{m}_{p}e^{-ip\theta} \quad ,\quad 
w^{n} (\theta) = \frac{1}{k} \sum_{p\in \mathbb{Z}}
w^{n}_{p}e^{-ip\theta} \ .
\end{equation}
In terms of Fourier modes, the constraints \eqref{AdSconditiononcomponents} and \eqref{AdSconditiongaugefixing} read
\begin{align}
\ell^{1}_{p} & \approx k\,\delta_{p,0} \, , & \ell^{0}_{p} &
\approx  0 \, , \nn \\
w^{2}_{p} & \approx 0 \, , & w^{1}_{p}&\approx 0 \, , \nn \\
w^{0}_{p} & \approx 0 \, , & w^{-1}_{p}&\approx 0 \, .
\end{align}
There are infinitely many of them and the matrix $C$ decomposes into
matrix blocks of infinite size,
\begin{equation}
C = \frac{1}{\sigma}\begin{pmatrix}
    0   &  2\sigma k \delta_{p+q,0} &  0  &  0  &  0  &  0  \\
-2 \sigma k\delta_{p+q,0} & 2ip\sigma k\delta_{p+q,0} & 0 & 0 & 0 & 0\\
0 & 0 & 0 & 0 & 0 & k \delta_{p+q,0}\\
0 & 0 & 0 & 0 & -3k\delta_{p+q,0} & ipk\delta_{p+q,0}\\
0 & 0 & 0 & 3k\delta_{p+q,0}& -\frac{3ipk}{2}\delta_{p+q,0}&-3\ell^{-1}_{p+q}\\
0 & 0 &-k\delta_{p+q,0}&-ipk\delta_{p+q,0}&3\ell^{-1}_{p+q} & 0
\end{pmatrix} .  
\end{equation}
To obtain this we need the brackets between the modes which we
collected in Appendix \ref{sec:formulae}. 
For instance, the second entry in the first row is determined from the block 
\be
\{\ell^{1}_{p},\ell^{0}_{q} \}=2\,\ell^{1}_{p+q}\approx
2k\,\delta_{p+q,0}
\ee
of eq.\ \eqref{fourier1}. The inverse of $C$ can be determined to be 
\begin{equation}
C^{-1} = \begin{pmatrix}
 -\frac{ip}{2k}\delta & -\frac{1}{2k}\delta & 0 & 0 & 0
& 0\\
\frac{1}{2k}\delta & 0 & 0 & 0 & 0 & 0\\
0 & 0 & \frac{-\sigma i
(p-q)}{k^{2}}\ell^{-1}_{-p-q}+\frac{\sigma ip^{3}}{6k}\delta & 
-\frac{\sigma}{k^{2}}\ell^{-1}_{-p-q} +\frac{\sigma p^{2}}{6k}\delta & 
-\frac{\sigma ip}{3k}\delta& -\frac{\sigma}{k}\delta\\
0 & 0 & \frac{\sigma}{k^{2}}\ell^{-1}_{-p-q} -\frac{\sigma p^{2}}{6k}\delta 
& \frac{\sigma ip}{6k}\delta & \frac{\sigma}{3k}\delta & 0\\
0 & 0& -\frac{\sigma ip}{3k}\delta & -\frac{\sigma}{3k}\delta & 0
& 0\\
0 & 0 & \frac{\sigma}{k}\delta & 0 & 0 & 0
\end{pmatrix}\ ,
\end{equation}
where $\delta$ stands for $\delta_{p+q,0}$.
From that we find the following induced Poisson structure, 
\begin{subequations}
\begin{align}
& i\{\ell^{-1}_{p},\ell^{-1}_{q} \}_{*} \,=\, (p-q) \ell^{-1}_{p+q} -
\frac{k}{2} p^{3}\delta_{p+q,0} \, , \\[5pt]
& i\{\ell^{-1}_{p},w^{-2}_{q} \}_{*} \,=\, (2p-q) w^{-2}_{p+q} \, , \\[2pt]
& i\{w^{-2}_{p},w^{-2}_{q} \}_{*} \,=\, \frac{kp^{5}}{96\sigma }\delta_{p+q,0} 
-\frac{1}{48\sigma } (p-q) (2p^{2}+2q^{2}-pq)\ell^{-1}_{p+q} \nn \\
& \phantom{i\{w^{-2}_{p},w^{-2}_{q} \}_{*} \,=\,}
+\frac{1}{3\sigma k} (p-q) \sum_{p'}\ell^{-1}_{p+q+p'}\ell^{-1}_{-p'} \,.
\end{align}
\end{subequations}
This is again the classical $\cW_3$-algebra with central charge
$c = 6k$. In fact, it can be related to~\eqref{W3fourier1} by identifying
\begin{equation}
\ell^{-1}_{p} \to - \cL_{-p}+\frac{k}{4}\delta_{p,0}\ , 
\qquad w^{-2}_{p} \to \frac{1}{4\sigma} \cW_{-p} \ .
\end{equation}

\subsection{Fall-off conditions for the metric-like fields}\label{sec:metric-like}

In Section \ref{sec:bound_cond} we identified asymptotically Anti-de Sitter solutions combining the condition \eqref{AdScondition} with \eqref{boundary1} and \eqref{gaugechoice}. We can now translate this into fall-off conditions for the metric-like fields $g_{\m\n}$ and $\vf_{\m\n\r}$. This allows for a direct comparison with the standard pure-gravity result of Brown and Henneaux \cite{BH} that further supports our choice. To this end we first have to express the metric-like fields in terms of the vielbein-like ones. The goal is the generalisation of the pure-gravity identity $g_{\m\n} = \h_{ab}\, e_\m{}^a\, e_\n{}^b$ and of the relation 
\eqref{metric_field} which is valid at the linearised level. The rationale behind both expressions is their invariance under local Lorentz-like gauge transformations. Therefore, we can look for their full non-linear analogues by imposing the invariance under the gauge transformations \eqref{gauge1} and \eqref{gauge2} generated by $\L^a$ and $\L^{ab}$. This fixes the structure of the metric-like fields up to a normalisation factor. For $\sigma = \tilde{\sigma}$ the result is
\begin{subequations}
\label{metriclikefields}
\begin{align}
& g \, = \,  e_a\, e^a \, - \, 2 \,\sigma \ e_{ab}\, e^{ab} \, , \label{metric_nonlin} \\
& \vf \, = \, e_a\, e_b\, e^{ab} \, + \, \frac{4}{3}\, \sigma \ e_{ac}\, e_b{}^c e^{ab} \, , \label{vf_nonlin}
\end{align}
\end{subequations}
where we omitted for brevity the form indices so that, for instance,
\be
g \, = \, \h_{ab} \left(\, e_\m{}^a\, e_\n{}^b - 2\, \sigma \, \h_{cd}\, e_\m{}^{ac} e_\n{}^{bd} \,\right) dx^\m \otimes dx^\n \, .
\ee
Notice that the definition of the metric receives a correction quadratic in the spin-3 vielbeins with respect to the pure-gravity expression. This is required by invariance under spin-3 Lorentz-like gauge transformations, while the two terms are independently invariant under the usual Lorentz transformations. The result crucially rests on the trace constraints. 

Eqs.\ \eqref{metriclikefields} admit a very convenient algebraic characterisation since they can be related to the quadratic and to the cubic Casimir of $sl(3)$, respectively. In fact, these expressions can be recovered from
\be \label{metric_casimir}
g_{\m\n} \, = \, \mathrm{tr} \left[\, e_{(\m}\cdot e_{\n)} \,\right] \, , \qquad\qquad \vf_{\m\n\r} \, = \, \frac{1}{9\sqrt{-\sigma}}\, \mathrm{tr} \left[\, e_{(\m} \cdot e_\n \cdot e_{\r)} \,\right] \, ,
\ee
where
\be
e_\m \, = \, e_\m{}^a \, J_a \, + \, e_\m{}^{ab} \, T_{ab}  \, .
\ee
The relation with the $sl(3)$ Casimir operators can be realised by noticing that, for any Lie algebra $\mathfrak{g}$ generated by $\{\cT_A\}$, a set of independent Casimir operators can be built as 
\be
C_p \, = \, a^{A_1 \ldots\, A_p} \, \cT_{A_1} \ldots \cT_{A_p} \, ,
\ee
where the fully symmetric $\mathfrak{g}$-invariant tensor $a^{A_1 \ldots\, A_p}$ is defined by
\be
a_{A_1 \ldots\, A_p} \,\sim\,\tr\left[\, \cT_{(A_1} \ldots \cT_{A_p)} \,\right] \, ,
\ee
and indices are lowered and raised with the Killing metric $\g_{AB}$ \cite{casimir}. The metric-like spin-3 field $\vf_{\m\n\r}$ is thus obtained from the contraction of the vielbeins with the symmetric rank-3 invariant tensor of $sl(3)$, in full analogy with the relation between the Riemannian metric $g_{\m\n}$ and the Killing metric.

Eqs.\ \eqref{metric_casimir} provide an intrinsic representation of the
metric-like fields. We can use them to express the metric-like fields
in terms of the unconstrained potentials $\hat{e},E$ associated to the
$\{L_i,W_m\}$ basis. Substituting
\be
e_\m \, = \, \hat{e}_\m{}^i \, L_i \, + \, E_\m{}^m \, W_m\, 
\ee
into eqs.\ \eqref{metric_casimir} we find for the metric 
\be \label{metricW}
g \, = \, - \, 4\, \hat{e}^{\,1} \hat{e}^{\,-1} + \, \hat{e}^{\,0} \hat{e}^{\,0} \, - \, \frac{4}{3}\,\sigma\, \left(\, 12\, E^2 E^{-2} \, - \, 3\, E^1 E^{-1} \, + \, E^0 E^0 \,\right) \, ,
\ee
and for the spin-3 field
\be \label{vfW}
\begin{split}
& \vf \, = \, \Big\{\, 4 \left(\, \hat{e}^{\,-1} \hat{e}^{\,-1} E^2 +\, \hat{e}^{\,1}\hat{e}^{\,1} E^{-2} \,\right) +\, \frac{4}{3}\, \hat{e}^{\,-1}\hat{e}^{\,1} E^0 + \, \frac{2}{3}\, \hat{e}^{\,0}\hat{e}^{\,0} E^0 \\
& \phantom{\vf \, = \, \Big\{} - \, 2\, \hat{e}^{\,0} \left(\, \hat{e}^{\,-1} E^1 + \, \hat{e}^{\,1} E^{-1} \,\right) \Big\} \\
& + \frac{4}{3}\,\sigma \, \Big\{\, 3 \left(\, E^{-1} E^{-1} E^2 + E^1 E^1 E^{-2} \,\right) - \, 8 \, E^{-2} E^0 E^2 - \, E^{-1} E^0 E^1 + \, \frac{2}{9}\, E^0 E^0 E^0 \,\Big\} \, .
\end{split}
\ee

In order to identify the fall-off conditions for the metric-like
fields one now has to substitute in eqs.\ \eqref{metricW} and
\eqref{vfW} the form of the general asymptotically AdS solution of the
field equations that was presented in \eqref{sol_gen} in terms of the
frame-like fields. Choosing for $\widetilde{A}$ the same normalisation
of eq.\ \eqref{sol_gen} but with a different overall sign, the result for the metric reads
\be \label{g_asym}
\begin{split}
g \, & = l^2 \frac{d r^2}{r^2} \, - \left\{\, r^2 \, + \, \left(8\pi G l\right)^2 \left(\frac{\cL(x^+)\widetilde{\cL}(x^-)}{r^2} + \frac{l^2}{4\sigma}\, \frac{\cW(x^+)\widetilde{\cW}(x^-)}{r^4} \right) \right\} dx^+ dx^- \\
& - \, 8\pi G l \left(\, \cL(x^+) (dx^+)^2 \, + \, \widetilde{\cL}(x^-) (dx^-)^2 \,\right) \, ,
\end{split}
\ee
where with respect to eq.\ \eqref{sol_einstein} we performed the change of coordinates $\r = \log (r/l)$. For the spin-3 field one obtains
\begin{align}
\vf \, & = \frac{l}{8\sigma} \,  (8\pi G l)   \, \left( \cW(x^+) (dx^+)^3 \, +
\, \widetilde{\cW}(x^-) (dx^-)^3 \right) \nn \\
& + \frac{l}{8\sigma} \, \left( 8\pi G l \right)^2 \, \left(\, 2 \,
\frac{\widetilde{\cL}(x^-)\cW(x^+)}{r^2} +  \left( 8\pi G l\right)\,
\frac{\cL(x^+)^2\widetilde{\cW}(x^-)}{r^4} \,\right) (dx^+)^2 dx^- \nn \\
& + \frac{l}{8\sigma} \, \left( 8\pi G l\right)^2 \,\left(\, 2 \, 
\frac{\cL(x^+)\widetilde{\cW}(x^-)}{r^2} +  \left( 8\pi G l\right)\,
\frac{\widetilde{\cL}(x^-)^2\cW(x^+)}{r^4} \,\right) (dx^-)^2 dx^+ \, .
\label{vf_asym}
\end{align}
Since eq.\ \eqref{sol_gen} solves eqs.\ \eqref{eq1} and \eqref{eq2},
these expressions provide exact solutions of their corresponding
second-order field equations, whose precise form is still to be
determined. However, it is also interesting to look at the leading
behaviour of eqs.\ \eqref{g_asym} and \eqref{vf_asym}. For instance, recognising that the metric appears in the Fefferman-Graham gauge \cite{FG}
\be
g \, = \, l^{2} \frac{d r^2}{r^2} \, + \, r^2 g_{ij}\, dx^i dx^j \, ,
\ee
one observes that at leading order the spatial metric satisfies the usual condition
\be \label{FG_asymptotics}
g_{ij} \, = \, \h_{ij} \, + \, \cO(1/r^2)
\ee
identifying an asymptotically $AdS_3$ solution with a flat boundary
metric. In this sense the boundary conditions of eq.\
\eqref{AdScondition} can be understood as those that do not spoil the
usual fall-off conditions for the metric identified long ago by Brown
and Henneaux \cite{BH}. In fact, the Fefferman-Graham asymptotic
conditions \eqref{FG_asymptotics} coincide with the Brown-Henneaux
ones up to residual boundary diffeomorphism. The full gauge fixing
leading to \eqref{AdSconditiongaugefixing} indeed implies that eqs.\ \eqref{g_asym} and \eqref{vf_asym} provide a fully gauge-fixed version of the admissible fall-off conditions (where the gauge fixing is meant to be performed with respect both to residual boundary diffeomorphisms and to residual boundary spin-3 gauge transformations). 

Notice also that the expressions for the metric-like fields can be cast in the form
\begin{align}
& g \, = \, l^{2}\frac{d r^2}{r^2} \, +\left(\, r^2\,\eta_{ij}\, dx^i dx^j \, - \, \cL_{ij}(x^m)\, dx^i dx^j \,\right) \, + \, \cO\left(r^{-2}\right) \, , \nn \\
& \vf \, = \, \cW_{ijk}(x^m)\, dx^i dx^j dx^k \, + \, \cO\left(r^{-2}\right) \, ,
\end{align}
where the tensors $\cL_{ij}$ and $\cW_{ijk}$ are traceless and conserved. In fact, in two-dimensions these conditions imply that they only have two independent components, one left-moving and one right-moving. In the boundary theory, which is defined on a flat background, these two objects are thus the Noether currents associated to the extended conformal symmetry we discovered in \eqref{W3}. 

\section{Comments on higher rank groups}\label{sec:generic_group}

In Section \ref{sec:cs_lin} we have shown how one can use $SL(n) \times SL(n)$ CS theories to describe a particular class of higher-spin interactions. The corresponding spectrum was identified simply by looking at what generators $T_{a_1 \ldots\, a_{s-1}}$ one needs in order to describe a $sl(n)$ algebra. To each $T_{a_1 \ldots\, a_{s-1}}$ one can then associate a spin-$s$ field. To this end, it is crucial to realise that the commutator \eqref{pre_JT} implies that the independent components of $T_{a_1 \ldots\, a_{s-1}}$ transform in the $(2s-1)$-dimensional irreducible representation under the adjoint action of $sl(2,\mathbb{R})$.
It is thus natural to associate a three-dimensional higher-spin
bosonic gauge theory to any CS theory based on a $G \times G$ gauge
group. The inclusion of fermions -- which we do not discuss in the
present paper -- will be obtained by considering also supergroups. The
selection of an embedding of $sl(2,\mathbb{R})$ in $\mathfrak{g}$ then
induces a branching of the generators in sets that transform
irreducibly under the adjoint action of $sl(2,\mathbb{R})$. This
should determine the spectrum. As in eq.\ \eqref{PM}, one could then associate the combinations
\be
\cP_A \, = \, \frac{1}{l} \left( \cT_A - \widetilde{\cT}_A \right) \, , \qquad\qquad \cM_A \, = \, \cT_A + \widetilde{\cT}_A 
\ee
of the generators $\{\cT_A\}$ and $\{\widetilde{\cT}_A\}$ of the two
copies of $\mathfrak{g}$ to vielbein-like and auxiliary fields,
respectively. This procedure was indeed proposed in a number of papers
dealing with higher-spin gauge fields in three space-time dimensions
\cite{general3D,pope,vasiliev_3d}.

However, in the $SL(n)\times SL(n)$ case we can go beyond this
identification and also control the elimination of auxiliary fields
and the recovering of the Fronsdal metric-like formulation. For this
reason, we shall mainly discuss the asymptotic symmetries emerging in
this class of examples, and in a generalisation that we shall describe
in a moment. First of all, notice that the procedure of Section
\ref{sec:cs_lin} selected not only the $sl(n)$ gauge algebra, but also
a particular embedding of $sl(2,\mathbb{R})$ in it: the principal
embedding. For a generic simple Lie algebra $\mathfrak{g}$, the
principal embedding has the property that the spins occurring in the
decomposition of $\mathfrak{g}$ into $sl (2,\mathbb{R})$
representations are $(l_i)$, where $l_i$, $i=1,\dots,r= {\rm
rank}(\mathfrak{g}$), are the exponents of the algebra \cite{Toda} and
$(l_i+1)$ are the ranks of the independent Casimir operators of
$\mathfrak{g}$. This suggests to extend the identification of
Section~\ref{sec:metric-like} between metric-like fields and
$\mathfrak{g}$-invariant tensors also to the $SL(n)\times SL(n)$
models of Section \ref{sec:cs_lin} and, in more generality, to all
higher-spin gauge theories obtained via the principal embedding of
$sl(2,\mathbb{R})$ into a simple $\mathfrak{g}$. Therefore, in the
following we shall focus on this class of higher-spin gauge theories,
which provide the most natural generalisation of the spin-3 example
that we discussed in detail in Section \ref{sec:symmetries}.

In order to discuss asymptotic symmetries in this context, let us
recall that if one imposes the boundary condition \eqref{boundary1},
then the boundary dynamics of the CS theory is described by a
Wess-Zumino-Witten model (see, for instance, the reviews
\cite{review_banados,carlip} and references therein). Furthermore, the
additional conditions \eqref{AdSconditiononcomponents} are those
inducing the Hamiltonian reduction of the WZW model to a $SL(3)$ Toda
theory \cite{WZW-Toda}. At this purely algebraic level, the reduction
of the affine algebra \eqref{affinealgebra} to the $\cW_3$-algebra
\eqref{W3fourier1} is often called Drinfeld-Sokolov (DS)
reduction. This is a general procedure that we can apply even beyond
the spin-3 case. Consider a generic $G\times G$ CS theory of the type
just discussed (i.e. characterised by the principal embedding of
$sl(2,\mathbb{R})$ into $\mathfrak{g}$). We can still fix the gauge as
in eq.\ \eqref{gaugechoice} and impose the natural condition
\begin{equation}\label{AdScondition2}
\left(\, A \, - \, A_{\text{AdS}} \,\right)\Big|_{\text{boundary}} = \, \mathcal{O} (1)\ .
\end{equation}
This extends the characterisation of asymptotically Anti-de Sitter
solutions in eq.\ \eqref{AdScondition} to the general case. We can
also impose the boundary condition \eqref{boundary1} and expand the
function $a(\theta)$ of eq.\ \eqref{solgaugefixed} as
\be
a(\theta) \, = \, \sum_{i=-1}^1 \ell^{\,i}(\theta)\, L_i \,+\, \sum_{i=2}^{r}\, 
\sum_{m=-l_i}^{l_i} \, w^{l_i,m}(\theta)\, W_{l_i,m} \, .
\ee
The first term is related to the lowest exponent $l_1$ which is always one. 
The $L_i$ are the $sl(2,\mathbb{R})$ generators while the $W_{l,m}$ generators are those transforming in the spin-$l$ representation under the 
adjoint action of 
$sl(2,\mathbb{R})$ (which one associates with spin-$(l+1)$ gauge fields).
Eq.\ \eqref{AdScondition2} then leads to the following constraints on the components of $a(\theta)$:
\be \label{generalconstraints}
\ell^{\,1} \,=\, 1 \, , \qquad w^{\,l,m} \,=\, 0 \qquad \forall\, l \ \ \textrm{and} \ \ \forall\, m > 0 \, .
\ee
We can use these first class constraints to reach the so called highest-weight gauge 
\cite{WZW-Toda} which is characterised by the additional conditions
\be
\ell^{\,0} \,=\, 0 \, , \qquad w^{\,l,m} \,=\, 0 \qquad \forall\, l \ \ \textrm{and} \ \ \forall\, m > -l+1 \, .
\ee
The remaining components $w^{l,-l}(\theta)$, which  are conformal primary fields of weight $l+1$ with respect to $\ell^{-1} \sim \cL$, enter $a(\theta)$ contracted with $W_{l,-l}$ generators. 
The asymptotic symmetries of these $G\times G$ CS theories are two copies of 
a $\cW$-algebra determined by $G$. In particular, for the 
$SL(n)\times SL(n)$ class we get two copies of the $\cW_n$-algebra.
In fact, with \eqref{generalconstraints}
we have recovered the constraints inducing the DS reduction of the affine Lie algebra identified in Section \ref{sec:boundary} \cite{WZW-Toda} and  
inspection of the results of \cite{WZW-Toda} suffices to arrive at this 
conclusion about the symmetry algebra.
The metric-like fields are presumably constructed as in Section 
\ref{sec:metric-like} and they are in 1-1 correspondence with the 
rank($\mathfrak{g}$) Casimir invariants on $\mathfrak{g}$. This 
is reminiscent of the generalised Sugawara construction in \cite{Bais}. 

The $\cW$-algebras which arise as asymptotic symmetry algebras have 
a central charge which, as we now demonstrate, has the same value as in the 
case of pure gravity.  
Recall that in eq.\ \eqref{sol_gen} we parameterised the space of asymptotically AdS solutions by simply adding the function $\cW(\theta)$ to the pure-gravity expression \eqref{BTZ}. It is therefore not surprising that we recovered the 
Brown-Henneaux value for the central charge in \eqref{W3}. A similar parameterisation of the space of asymptotically AdS solutions is obtained in the general case. The higher spin fields
do not modify the structure of the pure gravity part of $a(\theta)$ and therefore
the DS reduction leads to a $\cW$-algebra whose central charge 
has the Brown-Henneaux value.

In terms of CS data, the value of the central charge which arises in the 
DS reduction is (see eqs.\ (2.27) and (3.1) in~\cite{WZW-Toda})
\begin{equation}\label{centralcharge_general}
c \,=\, 12\,k \,\tr(L_0^2)\ .
\end{equation}
Comparing the Einstein-Hilbert action and the gravity sector
of \eqref{action} leads to the identification
\be
c\,=\,12\, k \, \tr(L_0^2) = \, \frac{3l}{2G}\, .
\ee

Eq.\ \eqref{centralcharge_general} holds also for DS reductions
performed with respect to different embeddings (see for instance the
classical, $k \to \infty$, limit of eq.\ (78) of \cite{non_principal}). But, as we already stressed, if we choose a different embedding we loose the suggestive correspondence between the spectrum of the theory and the Casimir operators of the underlying algebra
which  played a crucial role in Section \ref{sec:metric-like}.

\section{Conclusions}\label{sec:conclusions}

In this paper we studied the asymptotic symmetries of asymptotically
Anti-de Sitter solutions of higher-spin gauge theories coupled to
three-dimensional gravity with a negative cosmological constant. We
focussed on the case where only a finite number of bosonic higher-spin
fields is involved, for which we showed that the asymptotic symmetries
are described by two copies of a $\cW$-algebra selected by the
spectrum of the theory. These higher-spin models correspond to CS
theories based on a generic finite-dimensional $G \times G$ gauge
group. In particular, we discussed in detail the $SL(3) \times SL(3)$
example, describing the coupling of a spin-3 gauge field to
gravity. In this case we identified a $\cW_3 \otimes \cW_3$ algebra of
asymptotic symmetries. We also showed explicitly how to relate the
$SL(n)\times SL(n)$ CS theories to the standard frame-like
formulation of the higher-spin dynamics. Finally, we noticed that the
boundary conditions which select asymptotically AdS solutions in the
$SL(3)\times SL(3)$ example coincide with the constraints inducing the
Drinfeld-Sokolov reduction of a suitable $sl(3)$ affine Lie algebra to
$\cW_3$. Working in this framework we then discussed $G \times G$
higher-spin gauge theories based on simple Lie algebras $\mathfrak{g}$
where the gravitational sector is singled out by the principal
embedding of $sl(2,\mathbb{R})$ in $\mathfrak{g}$. In all cases the
value of the central charge of the resulting $\cW$-algebra is the same
as that found by Brown-Henneaux for pure gravity.

The choice of working with a finite number of higher-spin fields was
motivated by the simplicity of these models, that enabled us to
discuss in detail various aspects of the relation between CS theories
and higher-spin gauge theories (see for instance Section
\ref{sec:metric-like}). However, it will be interesting to also extend
our analysis to the three-dimensional higher-spin gauge theories of
\cite{blencowe,general3D}, that contain in their spectra an infinite
number of higher-spin gauge fields.\footnote{M.\ Henneaux and S.-J.\ Rey
have considered this case and they have constructed a non-linear
$W_\infty$ asymptotic symmetry algebra. We thank M.H.\ for informing
us of their results prior to publication of \cite{henneaux_rey}.}  In this respect they could
provide more realistic toy models for comparisons with the Vasiliev
theory \cite{vasiliev_int}, that describes an infinite tower of
interacting higher-spin gauge fields in AdS backgrounds with
$D\geq4$. Other directions that deserve further investigations are the
inclusion of fermions in the present framework and the study of CS
theories built upon other than the principal embedding of
$sl(2,\mathbb{R})$. While different embeddings were already discussed
in \cite{non_principal} in the framework of the Drinfeld-Sokolov
reduction, their interpretation as higher-spin gauge theories could
require some modifications with respect to the picture we have
presented.  The inclusion of Chan-Paton factors
\cite{deformed_oscillator} is also of interest, in particular in view
of a possible relation between higher-spin theories and open strings.
Another important aspect that requires further work is the
characterisation of the boundary theory. At the classical level the
Drinfeld-Sokolov reduction methods suggest that the relation between
three-dimensional gravity and Liouville theory
\cite{henneaux_vandriel} can be extended to a more general relation
between three-dimensional higher-spin gauge theories and Toda
theories.

\section*{Acknowledgements}

We are grateful to M.~Ba{\~n}ados, G.~Barnich, M.~Henneaux, O.~Mi\v{s}kovi\'{c},
S.-J.~Rey, \makebox{A.~Sagnotti} and M.~A.~Vasiliev for interesting discussions
and useful correspondence.

\begin{appendix}

\section{Conventions}\label{sec:conventions}

In this paper we adopt the mostly plus convention for the metric
\be
\h_{\,ab} \, = \, (\,-,+,+\,) \, ,
\ee 
while the Levi-Civita symbol is defined such that
\be
\e^{012} \, = \, - \ \e_{012} \, = \, 1 \, .
\ee

A pair of parentheses denotes the symmetrisation of the indices it encloses, with the minimum number of terms and without any normalisation factor. For instance, if $T_{ab}$ is a symmetric tensor
\be
V_{(a}T_{bc)} \, = \, V_a T_{bc} \, + \, V_b T_{ac} \, + \, V_c T_{ab} \, .
\ee
A vertical bar signals that the symmetrisation also encompasses the indices lying between the next bar and the closing parenthesis. For instance
\be
V_{(a|} W_d T_{|bc)} \, = \, V_a W_d T_{bc} \, + \, V_b W_d T_{ac} \, + \, V_c W_d T_{ab} \, .
\ee
In a similar fashion a pair of square brackets denotes the antisymmetrisation of the indices it encloses, again with the minimum number of terms and without any normalisation factor.

We normalise the Killing metric of $sl(3)$ such that in the basis
$\{\cT_A\} = \{L_{i},W_{m}\}$ with $i=-1,\ldots,1$, $m=-2,\ldots,2$
(see eq.\ \eqref{sl3can}), it is given by 
\begin{equation} \label{killingmetric}
\gamma_{AB} \, = \, \begin{pmatrix}
 0 & 0 &-1 & 0 & 0  & 0 & 0   & 0 \\
 0 &1/2& 0 & 0 & 0  & 0 & 0   & 0 \\
-1 & 0 & 0 & 0 & 0  & 0 & 0   & 0 \\
 0 & 0 & 0 & 0 & 0  & 0 & 0   & -4\sigma \\
 0 & 0 & 0 & 0 & 0  & 0 & \sigma & 0 \\
 0 & 0 & 0 & 0 & 0  &-\frac{2}{3}\sigma& 0   & 0 \\
 0 & 0 & 0 & 0 & \sigma & 0 & 0   & 0 \\
 0 & 0 & 0 & -4\sigma & 0  & 0 & 0   & 0
\end{pmatrix}\ .
\end{equation}
The invariant form which is used to write the CS action \eqref{action} is 
defined as
\be\label{trAB}
\tr(\cT_A\cT_B) \, = \, \g_{AB} \, .
\ee
For the $sl(2)$ generators in the basis $J_a$ this reproduces
\eqref{convention_killing}.  This convention preserves the standard
pure-gravity relation \eqref{level} between the level $k$ of the CS
theory and the cosmological constant.

A possible $3 \times 3$ matrix realisation of the
algebra \eqref{sl3can} is given by the fundamental representation of $sl (3)$,
\begin{alignat}{8}
& L_1 \, = \, 
\begin{pmatrix}
0 & 0 & 0 \\
1 & 0 & 0 \\
0 & 1 & 0	
\end{pmatrix} , \qquad\qquad
& & L_{-1} \, = \, 
\begin{pmatrix}
0 & -2 & 0 \\
0 & 0 & -2 \\
0 & 0 & 0	
\end{pmatrix} , \qquad\qquad \nn \\
& L_0 \, = \, 
\begin{pmatrix}
1 & 0 & 0 \\
0 & 0 & 0 \\
0 & 0 & -1	
\end{pmatrix} , \qquad\qquad
& & W_{0} \, = \, \frac{2}{3} \sqrt{-\sigma}
\begin{pmatrix}
1 & 0 & 0 \\
0 & -2 & 0 \\
0 & 0 & 1	
\end{pmatrix} , \nn \\
& W_{1} \, = \, \sqrt{-\sigma}
\begin{pmatrix}
0 & 0 & 0 \\
1 & 0 & 0 \\
0 & -1 & 0	
\end{pmatrix} \, , \qquad\qquad
& & W_{-1} \, = \, \sqrt{-\sigma}
\begin{pmatrix}
0 & -2 & 0 \\
0 & 0 & 2 \\
0 & 0 & 0	
\end{pmatrix} , \nn \\
& W_{2} \, = \, 2\, \sqrt{-\sigma}
\begin{pmatrix}
0 & 0 & 0 \\
0 & 0 & 0 \\
1 & 0 & 0	
\end{pmatrix} , \qquad\qquad
& & W_{-2} \, = \, 2\, \sqrt{-\sigma}
\begin{pmatrix}
0 & 0 & 4 \\
0 & 0 & 0 \\
0 & 0 & 0	
\end{pmatrix} . \label{matrix_rep}
\end{alignat}
Notice that the representatives of the $W_m$ generators are real for
$\sigma < 0$, in agreement with the association of the real form
$sl(3,\mathbb{R})$ with negative values of $\sigma$. Furthermore,
comparison with \eqref{killingmetric} reveals that in the fundamental
representation `$\tr$' in \eqref{trAB} denotes one quarter times the
matrix trace.

\section{Useful formulae}\label{sec:formulae}

The non-vanishing Dirac brackets \eqref{affinealgebra} for the $sl(3)$ modes \eqref{modesW} read
\begin{align}
\{\ell^{1}_{p},\ell^{0}_{q} \} & = 2\,\ell^{1}_{p+q} &
\{\ell^{1}_{p},\ell^{-1}_{q} \} & = \ell^{0}_{p+q} - ipk\,\delta_{p+q,0} \nn \\
\{\ell^{0}_{p},\ell^{-1}_{q} \} &= 2\,\ell^{-1}_{p+q} &
\{\ell^{0}_{p},\ell^{0}_{q} \} & = 2\,ipk\,\delta_{p+q,0} \label{fourier1}
\end{align}
\begin{align}
\{w^{1}_{p},\ell^{1}_{q} \} & = -\,4\, w^{2}_{p+q} 
& \{w^{2}_{p},\ell^{0}_{q} \} & = 4\, w^{2}_{p+q} 
& \{w^{2}_{p},\ell^{-1}_{q} \} & = w^{1}_{p+q} \nn \\
\{w^{0}_{p},\ell^{1}_{q} \} & = -\,3\, w^{1}_{p+q} 
& \{w^{1}_{p},\ell^{0}_{q} \} & = 2\, w^{1}_{p+q} 
& \{w^{1}_{p},\ell^{-1}_{q} \} & = 2\, w^{0}_{p+q} \nn \\
\{w^{-1}_{p},\ell^{1}_{q} \} & = -\,2\, w^{0}_{p+q} 
& \{w^{-1}_{p},\ell^{0}_{q} \} & = -\,2\, w^{-1}_{p+q}  
& \{w^{0}_{p},\ell^{-1}_{q} \} & = 3\, w^{-1}_{p+q} \nn \\
\{w^{-2}_{p},\ell^{1}_{q} \} & = -\, w^{-1}_{p+q} 
& \{w^{-2}_{p},\ell^{0}_{q} \} & = -\,4\, w^{-2}_{p+q} 
& \{w^{-1}_{p},\ell^{-1}_{q} \} & = 4\, w^{-2}_{p+q} \label{fourier2}
\end{align}
\begin{align}
\{w^{2}_{p},w^{-1}_{q} \} & = \frac{1}{\sigma}\, \ell^{1}_{p+q} &
\{w^{2}_{p},w^{-2}_{q} \} & = \frac{1}{2\sigma}\, \ell^{0}_{p+q} - \frac{ipk}{4\sigma}\, \delta_{p+q,0} \nn \\
\{w^{1}_{p},w^{0}_{q} \} & = -\, \frac{3}{\sigma}\, \ell^{1}_{p+q} & 
\{w^{1}_{p},w^{-1}_{q} \} & = -\, \frac{1}{\sigma}\, \ell^{0}_{p+q} + \frac{ipk}{\sigma}\, \delta_{p+q,0} \nn \\
\{w^{-1}_{p},w^{0}_{q} \} & = \frac{3}{\sigma}\, \ell^{-1}_{p+q} &
\{w^{0}_{p},w^{0}_{q} \} & = -\, \frac{3ipk}{2\sigma}\, \delta_{p+q,0} \nn \\
\{w^{-2}_{p},w^{1}_{q} \} & = -\, \frac{1}{\sigma}\, \ell^{-1}_{p+q} \label{fourier3}
\end{align}

\end{appendix}


\end{document}